\documentclass{article}
\usepackage{geometry}
\usepackage{graphicx}
\usepackage{subfigure}
\usepackage{amssymb}
\usepackage{amsmath}
\usepackage{epstopdf}
\usepackage{cite}
\usepackage{bbm}
\usepackage{multirow}
\usepackage{enumerate}
\usepackage{ulem}
\usepackage{xcolor}
\usepackage{appendix}
\usepackage[hyperindex=true,
          pdfstartview=FitH,
          bookmarksnumbered=true,
          bookmarksopen=true,
          citecolor=blue,
          linkcolor=blue,
          colorlinks=true,
          pdfborder=00]{hyperref}

\newcommand{\ba}{\begin{align}}
\newcommand{\ea}{\end{align}}
\newcommand{\be}{\begin{equation}}
\newcommand{\en}{\end{equation}}
\newcommand{\bea}{\begin{eqnarray}}
\newcommand{\ena}{\end{eqnarray}}

\usepackage{graphicx} % Required for inserting images

\title{Observability of gravitational waves excited by binary stars orbiting around a supermassive black hole by space-based gravitational wave observatory}
\author{Kun Meng$^1$\footnote{kunmeng@wfu.edu.cn},  Hongsheng Zhang$^2$\footnote{sps\_zhanghs@ujn.edu.cn},  Xi-Long Fan$^3$\footnote{xilong.fan@whu.edu.cn},  Yong Yuan$^3$, Fei Du$^1$
\footnote{mhdeng@wfu.edu.cn}\\
$^1$School of Physics and Electronic Information,
	Weifang University, Weifang 261061, China \\
$^2$School of Physics and Technology, University of Jinan, Jinan, China\\
$^3$School of Physics and Technology, Wuhan University, Wuhan, Hubei 430072, China
}

\date{\today}

\begin{document}

\maketitle

\begin{abstract}
We produce the gravitational waveforms for the extreme mass ratio inspiral systems (EMRIs) of binary stars moving around central supermassive black hole (SBH), or called B-EMRIs.
  We calculate the \textcolor{black}{external} orbits of \textcolor{black}{the binary stars} via the \textcolor{black}{commonly used} Hamilton-Jacobi \textcolor{black}{(HJ)} approach, \textcolor{black}{and calculate the internal orbits of the binary stars via Lagrangican approach}. To improve accuracy we adopt the quadrupole-octupole expression of gravitational wave (GW) and
  study the contribution of radiation reaction. Compared to the waveforms of EMRIs, there are higher frequency oscillations superposed on the waveforms of B-EMRIs. We perform frequency spectrum
  analysis of the GW waveforms, and find that higher frequency signals give their prominency in the waveforms of B-EMRIs. To obtain high precise result for future observation of GWs from
  space-based detector, we take into account gravito-electromagnetic (GEM) force, and compare the waveforms of B-EMRIs with \textcolor{black}{GEM effects against those of B-EMRIs without GEM effects and against those }  of EMRIs. The result of mismatch shows that the waveforms of
  B-EMRIs are credibly distinguishable by the space-based GW detectors  when GEM force is considered.
\end{abstract}

\section{Introduction}
\label{section1}

Black hole (BH) may be the most thoroughly-studied object before its discovery in the history of science. Recently, the Event Horizon Telescope (EHT) collaboration released images of the supermassive BHs M87* and SgrA*\cite{EventHorizonTelescope:2019ggy,EventHorizonTelescope:2022xqj}, while ground-based GW detectors have reported multiple detections of GW signals\cite{LIGOScientific:2016aoc,LIGOScientific:2016sjg}. In addition, recent observations of ringdown signals from binary mergers indicate that the properties of the remnants align with the Kerr metric\cite{KAGRA:2025oiz}. All these observations confirm the existence of black holes, and launch a new era of testing gravity in strong-field regime and far beyond the scale of solar system. 

One of the primary objectives of space-based gravitational wave detectors is to capture gravitational waves emitted by EMRIs. 
EMRIs are composed of a central SBH and stellar-mass compact objects moving around it. The multi-body interaction in the cusp of the stellar population surrounding the SBH may scatter the compact objects into the orbits close the central SBH and form the EMRIs\cite{Sigurdsson:1997vc,Sigurdsson:1996uz}. The gravitational waves (GWs) generated by EMRIs encode information about both the central SBH and the orbiting stellar-mass compact object. Therefore, their detection will enable us to achieve key scientific objectives, including measuring black hole mass and spin, probing their strong-field properties and multipole structure, determining the nature of the inspiraling compact object, tracing galactic evolutionary history, and constraining the expansion rate of the universe. These observations will provide experimental evidence to advance our understanding of the fundamental theory of gravity and the cosmos.

The waveforms of GWs are known to be critical for data analysis to extract the information of black holes. The waveforms generated by motions of single stellar-mass compact object in BH spacetime have been studied
intensively\cite{Hughes:2001jr,Glampedakis:2002ya,Gupta:2022fbe,Polcar:2022bwv,Munna:2022xts,
Kerachian:2023oiw,Takahashi:2023flk,Burko:2013cca,Glampedakis:2005hs,Babak:2017tow,Munna:2019fjz,Zi:2021pdp,Destounis:2022obl,Zi:2023pvl,Barack:2003fp,
Zi:2023omh,Dai:2023cft,Pan:2023wau}, including probe of fundamental fields\cite{Zhang:2022rfr,Liang:2022gdk}, modified
gravity\cite{Will:1977wq,Will:1989sk,Barausse:2015wia,Xu:2021kfh,Tan:2024utr,Gair:2011ym}, extra dimensions\cite{Rahman:2022fay}, electric charge and internal structure of
small body objects\cite{Zi:2022hcc,Zhang:2022hbt,Yang:2022tma,Drummond:2023loz,Mino:1995fm,Toshmatov:2020wky,Piovano:2020zin}, etc. Binary-star systems are important
astrophysical objects which are constituted of binary stars revolving around each other. The amount of binary-star systems is huge in the Galaxy. It's estimated that the
amount of binary-star systems is no less than that of single star.  The vast majority of stars are located in a binary system. It is estimated that more than  70\% of all
massive stars may exist in systems with two or three stars.

 In active galactic nuclei, due to the rich forms of interactions between SBH, the gaseous accretion disks and the stellar mass compact objects, the binary may be formed and may be put onto the orbit closed to the central SBH. There are several mechanisms to produce binary stars orbiting close to a SBH, such as the migration trap by accretion disc dynamics \cite{Bellovary:2015ifg,Secunda:2020mhd} and tidal capture processes \cite{Chen:2018axp,Generozov:2018niv}. Since the binary stars are considerable popular in galaxies, they may be captured by the SBH centered at the galaxies. If the resulting binary is stable against tidal disruption from the SBH \cite{Hills:1988cvl,Suzuki:2020vfw}, a hierarchical triple is formed \cite{Han:2018hby,Addison:2015bpa,Naoz:2011eic,Naoz:2011mb}. Chen and Han first pointed out that B-EMRI systems can form through the process of tidal capture\cite{Chen:2018axp}. With the parameters selected in this paper, calculation shows the distance between the supermassive black hole and the binary system is significantly greater than the tidal disruption radius, so the binary system will not be disrupted by tidal forces, a stable triple system can be formed. Notably, a binary system of two stellar-mass objects has recently been detected in close orbit around Sgr A* \cite{Peissker:2024ade}. The binary stars orbiting around central SBH constitute a significant typical EMIRs. Thus, it is quite sensible to
consider the waveforms produced by motions of binary stars surrounding a SBH in addition to the case of single star. The GW generated by each of the binary star
overlaps, so the waveforms generated by this type of EMRIs are distinguishable from that generated by single star. In this paper, we  consider the case that the amplitudes
of the GW generated by internal motion of binary stars are small compared to that generated by external motion of the mass center, so it's expected small perturbations would
appear in B-EMRIs when compared to the waveforms of ordinary EMRIs. It's really worthwhile  to study GW waveforms of such systems  for future data analysis of space-based detectors for GWs. The GWs emitted from B-EMRIs have recently been studied in \cite{Jiang:2024mdl,Yin:2024nyz}, some interesting effects such as resonant excitation and gravitational lens have been investigated  in \cite{Santos:2025ass,Santos:2026lzq}.

The objective of this paper is to distinguish EMRIs from B-EMRIs via gravitational waves at the level of a coarse approximation, enabling the differentiation of waveforms observed by future gravitational wave detectors. We focus on the primary distinguishing feature between EMRIs and B-EMRIs, employing the lowest-order approximation and neglecting more refined effects. For simplicity, we omit perturbations to the inner orbits---such as Kozai-Lidov oscillations \cite{Antonini:2009dj,Bradnick:2017pww} induced by the tidal forces of the SBH---as well as the inspiral of the inner binary itself. As we will see below, under such approximation the waveforms exhibit clear oscillations for binary star systems compared to those generated by a single star orbiting the SBH.
 In this paper, we adopt the ``numerical kludge''
(NK) method which generates waveforms more conveniently and quickly compared to the Teukolsky method \cite{Drasco:2005kz,Hughes:1999bq,Cutler:1994pb,Tanaka:1993pu,Shibata93,Tagoshi:1995sh,Poisson:1993vp,Cutler:1993vq,Apostolatos:1993nu,
Poisson:1993zr,Poisson:1994yf,Tagoshi:1994sm,Shibata:1994jx}, this method are proved to be feasible, it captures the principle feature of true waveforms\cite{Babak:2006uv} so it meets our needs. To improve accuracy, aside from the mass quadrupole we take
into account the  mass octupole and current quadrupole moments of the source to solve the wave equations. It's justified the NK method does excellently at approximating the
true gravitational waveforms so this method has already been used for scoping out data analysis of LISA.
For the radiation reaction, we consider the adiabatic evolution of orbit parameters: the timescale of evolution of the orbit parameters is much larger than that of orbit
cycles such that the motion of mass point in BH spacetime is nearly geodesic. Further, since the amplitude of GWs generated by internal motions are smaller compared to that
generated by external motions of the mass center, to master the principle property of the waveform we concentrate on the radiation reaction due to the GW emissions generated by the motion of mass
center. We take the hybrid scheme that combines Tagoshi's\cite{Tagoshi:1995sh}, Shibata's\cite{Shibata:1994jx} and Ryan's fluxes\cite{Ryan:1995zm,Ryan:1995xi}, whose precision is comparable or higher than
Teukolsky-based results for
generic circular-inclined orbits \cite{Gair:2005ih}. 

According to equivalence principle only at the origin of free-fall frame (FFF), i.e., the mass center of the binary, the spacetime is flat, if there is departure from the origin the effects of
curved spacetime appear. Since for binary system, either star does not locate at the origin of FFF,  we should take into account the gravitational effects caused by this reason. Later we'll study
the effects of GEM force on the waveforms of B-EMRIs.

The paper is organised as follows. In section \ref{section2} we calculate the trajectories of the binary stars and give waveforms of GW with the NK method. In section
\ref{section3} we analyse the waveforms by calculating mismatch between the waveforms of binary stars and that of single star. We summarize our results in the last section.
In this work, we adopt the geometrized units with $G=c=1$.

\section{Numerical Kludge Waveforms\label{section2}}
 In this section we adopt the NK waveforms approach developed
  in \cite{Babak:2006uv} to produce the GW waveforms of the B-EMRIs. We consider the system as, the central SBH is Kerr, the stellar-mass binary stars revolve
  around each other and they as a whole move along the inspiral trajectory of the central SBH. The binary stars are supposed to live far away from the SBH such that the internal gravity
  between them dominates, so the impacts of the SBH on the internal orbit of the binary stars are neglected, while the center of mass of the binary stars is still on the
  inspiral trajectory of the SBH. The geometry of binary stars moving in the spacetime of SBH is illustrated in Fig.\ref{emri}.

\subsection{the geometry  of the orbit}

\begin{figure}[h]
\begin{center}
\includegraphics[width=0.9 \textwidth]{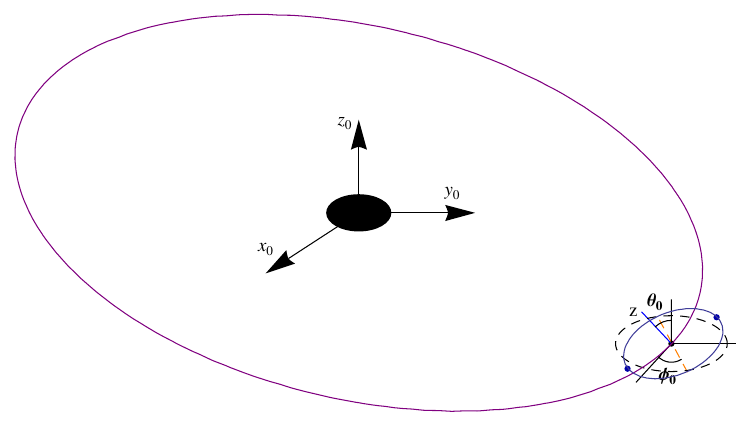}
\end{center}
\caption{A sketch of binary stars orbiting around the central SBH, in which the internal orbit is exaggerated to clearly show the positional relation}\label{emri}.
\end{figure}

The binary are supposed to revolve around each other along  elliptic orbit. We introduce the relative coordinates $\vec{\tilde{r}}=\vec{r}_1-\vec{r}_2$, with
$\tilde{r}=\frac{\tilde{p}}{1+\tilde{e}\cos\tilde{\beta}}$, then the coordinates of the binary with masses $m_1$ and $m_2$ can be expressed with coordinates of mass center and
relative coordinates as $\vec{r}_1=\vec{r}_C+\vec{\tilde{r}}_1, \vec{r}_2=\vec{r}_C+\vec{\tilde{r}}_2$, where $\vec{\tilde{r}}_1=\frac{m_2}{m_1+m_2}\vec{\tilde{r}},
\;\vec{\tilde{r}}_2=-\frac{m_1}{m_1+m_2}\vec{\tilde{r}}$. We use the coordinates and parameters with tilde to describe the internal motion of binary stars around each other.
The spin axis of the central SBH is along $z_0$, as displayed in Fig.\ref{emri}, $\theta_0$ is the angle between $z_0$ and $z$ which is the normal direction of the surface of binary stars' internal
elliptic
orbit. $\phi_0$ is the angle between $x_0$ and the node line which is the intersection line between the surface of binary stars' internal orbit and the horizonal surface.

Now let's discuss the orbit of a mass point in Kerr spacetime, in Boyer-Lindquist coordinates Kerr metric can be expressed as
\begin{equation}
ds^2=-\left(1-\frac{2M}{\Sigma}\right)dt^2+\frac{\Sigma}{\Delta}dr^2-\frac{4a Mr\sin^2\theta}{\Sigma}dtd\phi+\Sigma
d\theta^2+\sin^2\theta\left(r^2+a^2+a^2\sin^2\theta\frac{2Mr}{\Sigma}\right)d\phi^2,
\end{equation}
where $\Delta=r^2-2Mr+a^2$ and $\Sigma=r^2+a^2\cos^2\theta$.
 \textcolor{black}{In} the  \textcolor{black}{HJ} formulation, the \textcolor{black}{motion of a mass point is governed by}  \textcolor{black}{HJ} equation \textcolor{black}{which} reads
\begin{equation}
-\frac{\partial S}{\partial\tau}=\frac{1}{2}g^{\alpha\beta}\frac{\partial S}{\partial x^\alpha}\frac{\partial S}{\partial x^\beta}.\label{HJeq}
\end{equation}
The Kerr BH admits two obvious Killing vectors $\partial_t$ and $\partial_\phi$, which correspond to two conserved charges, energy $E$ and angular momentum $L_z$, so for the mass point with mass
$\mu$ ($\mu= m_1 +m_2$ in this paper), the action reads
\begin{equation}
S=\frac{1}{2}\mu^2\tau-Et+L_z\phi+S_r(r)+S_\theta(\theta),\label{HJaction}
\end{equation}
which leads to $p_r=dS_r/dr$ and $p_\theta=dS_\theta/d\theta$. Inserting (\ref{HJaction}) into the Hamilton-Jacobi equation (\ref{HJeq}) and separating variables one obtains the specific forms of $S_r$ and $S_\theta$. 
The separation constant $C$ is related with Carter constant $Q$ through
$Q=C-(L_z- aE)^2$. Actually, the variable-separability of  \textcolor{black}{HJ} equation in Kerr spacetime corresponds to a symmetry which is described by a Killing tensor
which satisfies $\nabla_{(\lambda}K_{\mu\nu)}=0$. This nontrivial symmetry corresponds to the third conserved quantity called Carter constant \cite{Carter:1971zc}.
The explicit integration expression of coordinates can be obtained by setting the partial derivative of \textcolor{black}{HJ} function with respect to the constants
of motion to be zero, that's to say the motions of particles in Kerr spacetime are integrable.

\begin{figure*}[h]
\begin{center}
\includegraphics[width=0.23 \textwidth]{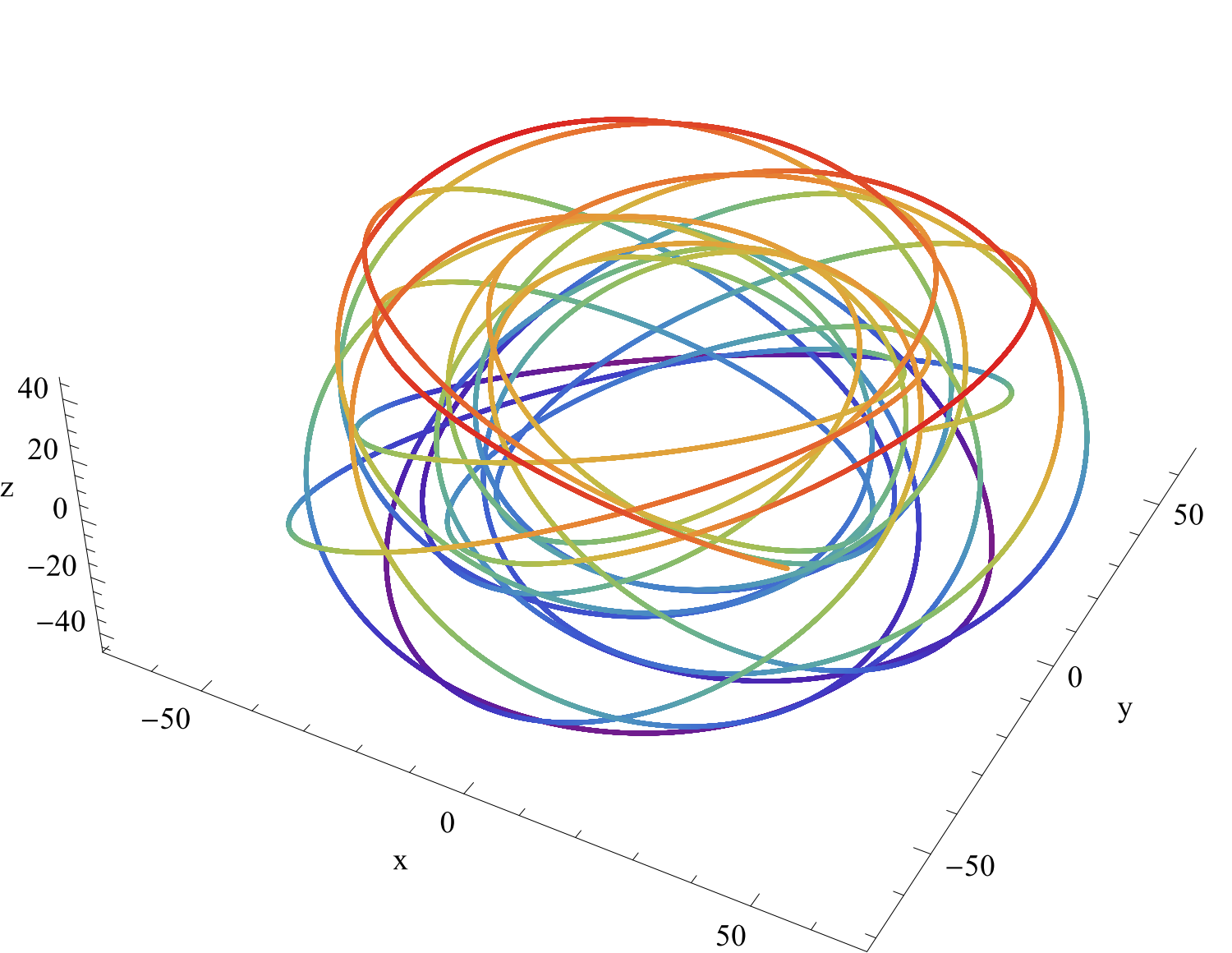}
\includegraphics[width=0.23 \textwidth]{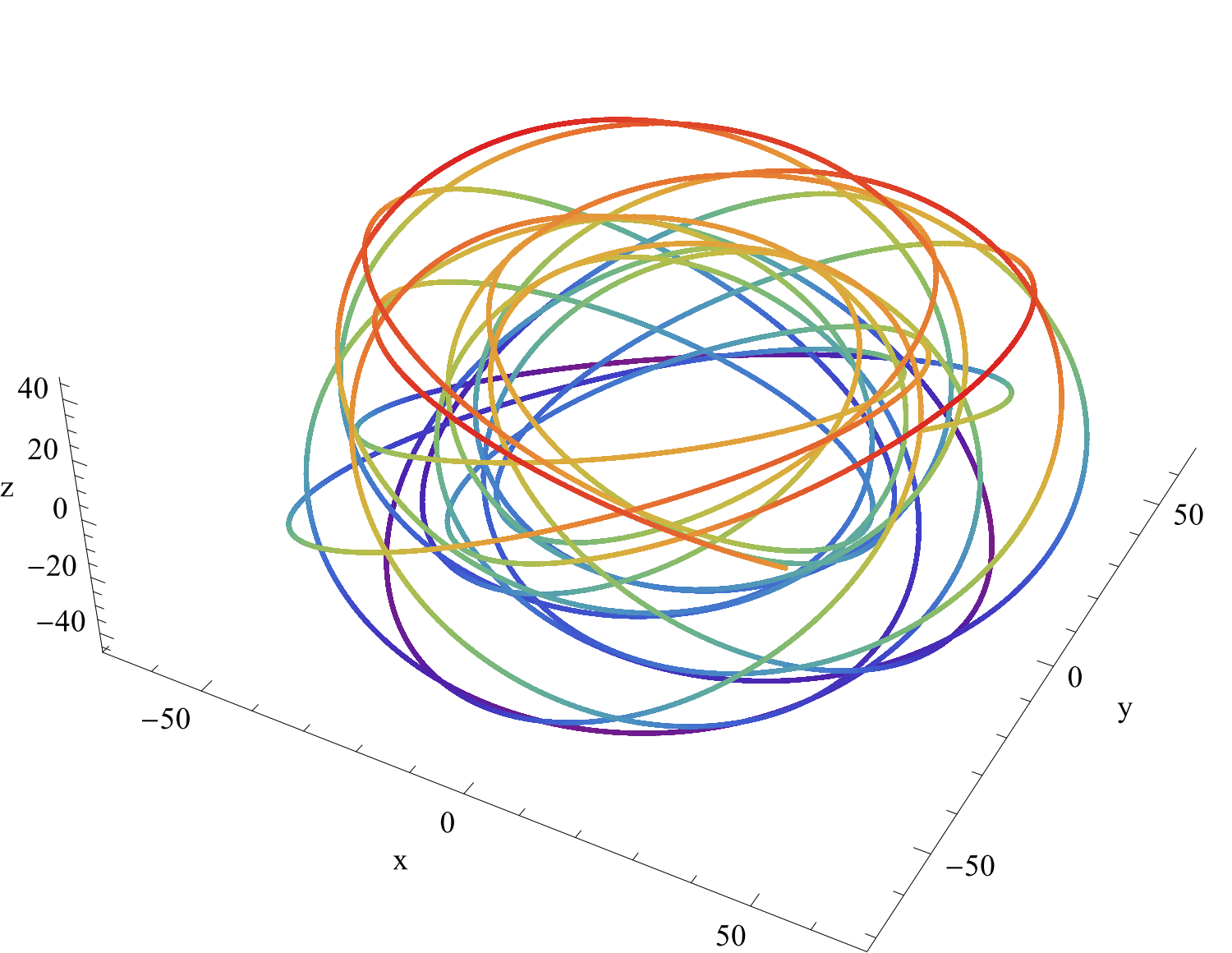}
\includegraphics[width=0.23 \textwidth]{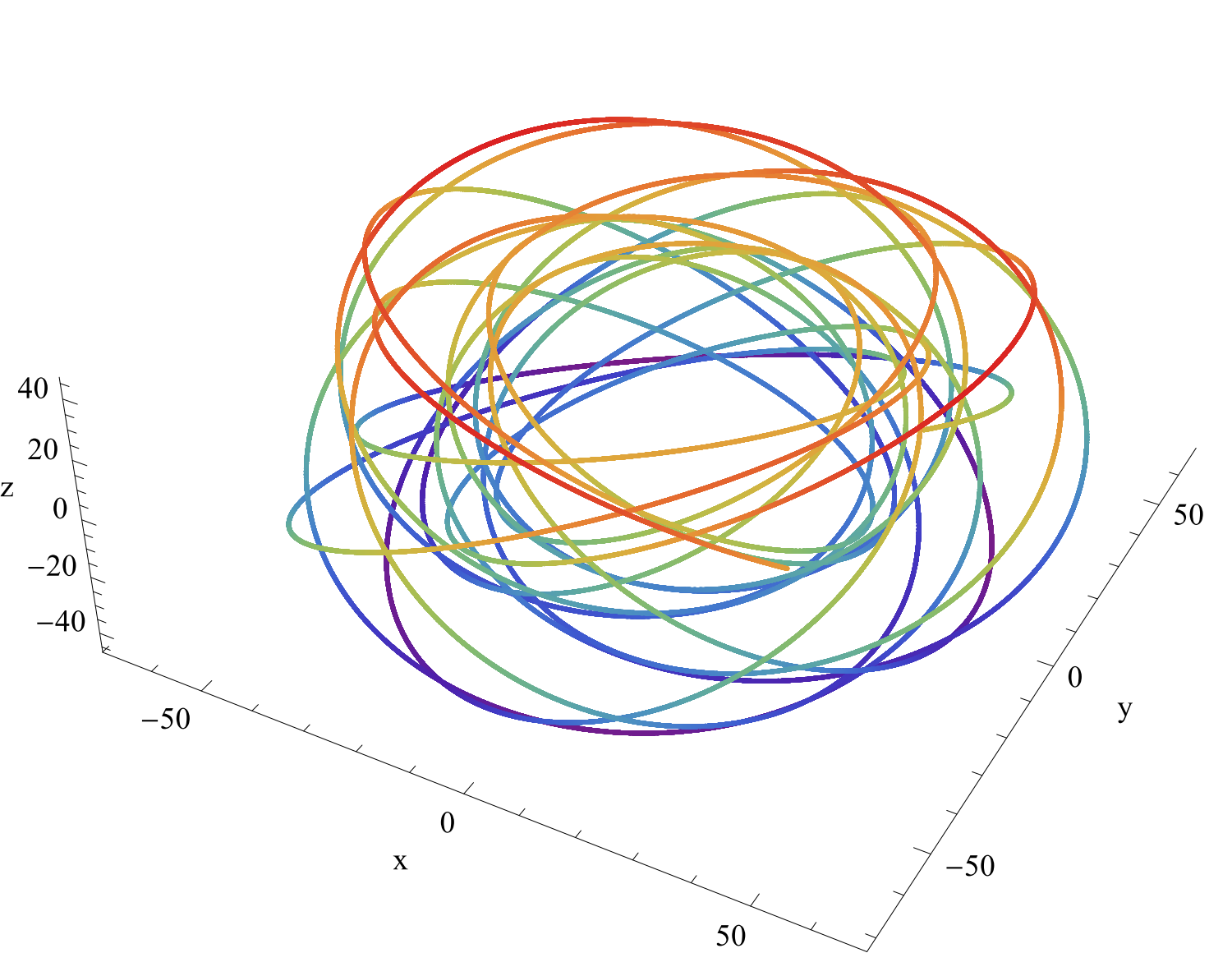}
\includegraphics[width=0.23 \textwidth]{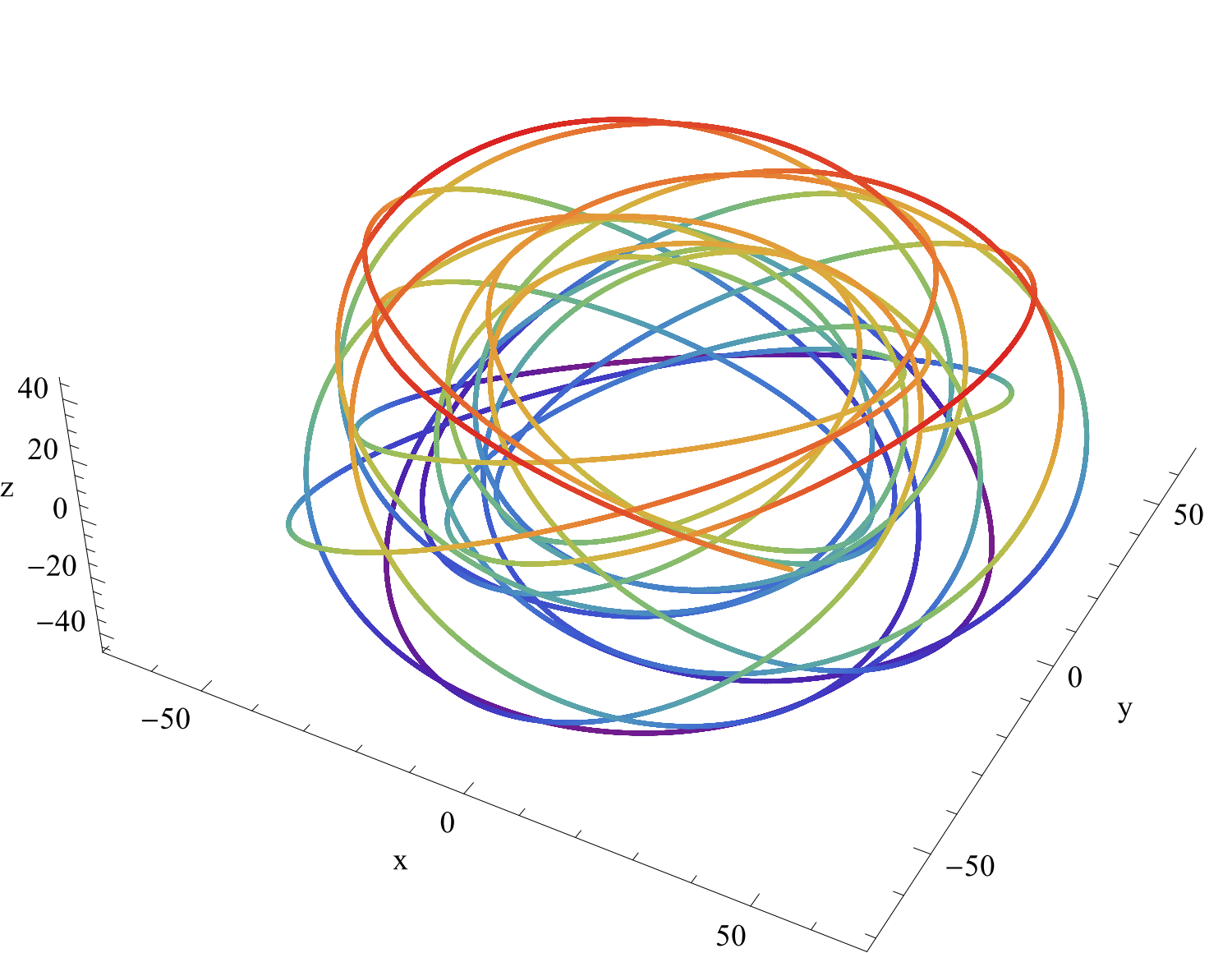}
\includegraphics[width=0.23 \textwidth]{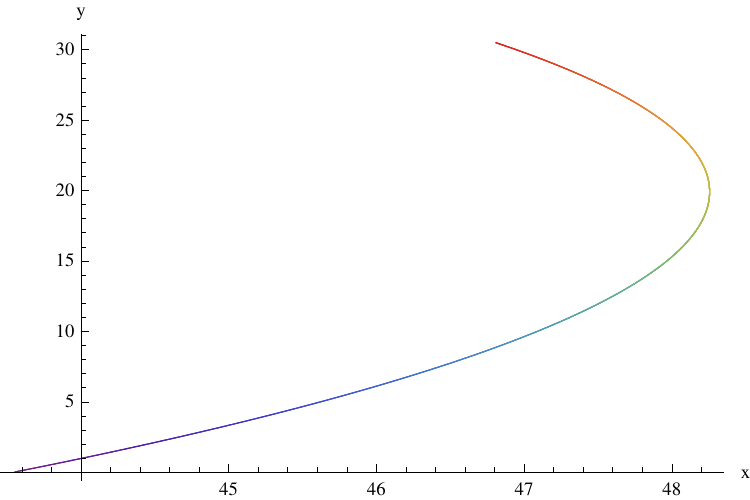}
\includegraphics[width=0.23 \textwidth]{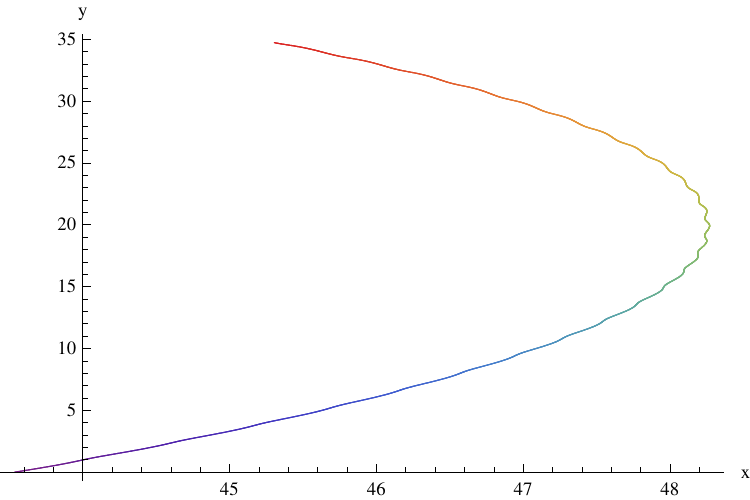}
\includegraphics[width=0.23 \textwidth]{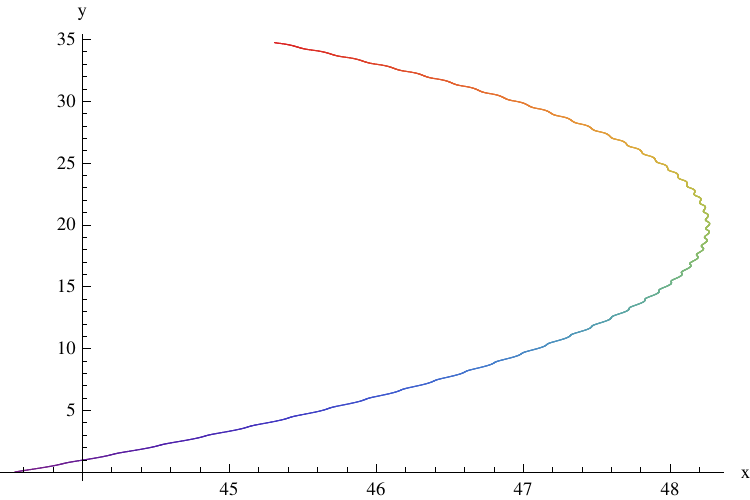}
\includegraphics[width=0.23 \textwidth]{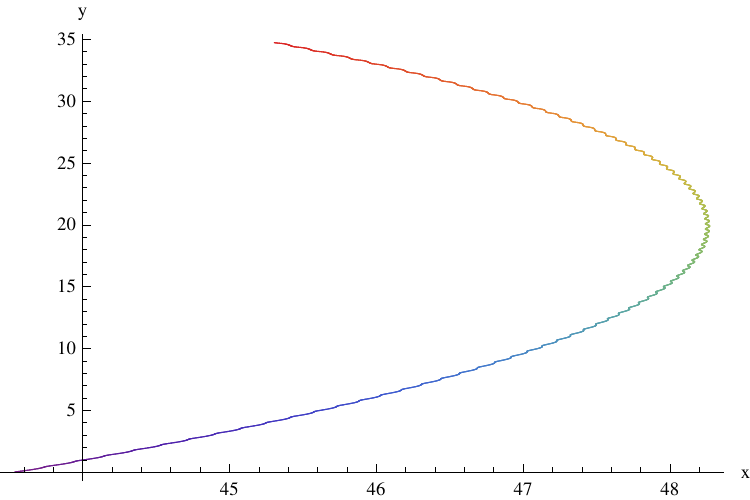}
\end{center}
\caption{The first panel displays the orbit of compact object in EMRIs, the second to fourth panels display the orbits of one of the compact objects in B-EMRIs. The lower four panels display the
details of the corresponding orbits in the upper four panels in the x-y plane. For the first panel, the mass of the single compact object is set to be $m=18M_{\odot}$, the initial external orbit
parameters are set to be $p(0)=10M, e(0)=0.3, \iota(0)=7\pi/9$. For the second to fourth panels, the masses of the compact objects are set to be $m_1=2M_{\odot}, m_2=3M_{\odot}$; $m_1=8M_{\odot},
m_2=10M_{\odot}$ and $m_1=20M_{\odot}, m_2=30M_{\odot}$ respectively. The initial external orbit parameters are set to be $p(0)=10M, e(0)=0.3$ as well,  the internal orbit parameters are set to be
$\tilde{p}=0.005M, \tilde{e}=0.2$. The rotation parameter of the SBH is set to be $a=0.9M$.}\label{orbit}
\end{figure*}

With the above results, one now is able to give the geodesics of a mass point in Kerr spacetime
\begin{equation}
\begin{aligned}
\Sigma \frac{dr}{d\tau} &= \pm \sqrt{V_r},\\
\Sigma \frac{d\theta}{d\tau} &= \pm \sqrt{V_{\theta}},\\
\Sigma \frac{d\phi}{d\tau} &=  V_\phi, \\
\Sigma \frac{dt}{d\tau} &=  V_t,\label{geodesic}
\end{aligned}
\end{equation}
with
\begin{equation}
\begin{aligned}
V_r &= \left[ E(r^2 + a^2) - L_z a \right]^2 -
\Delta\left[\mu^2 r^2 + (L_z-aE)^2 + Q\right],\\
V_{\theta} &=  Q -\cos^2{\theta}\left[  a^2(\mu^2-E^2)
+ \frac{L_z^2}{\sin^2{\theta}} \right],\\
V_{\phi} &= \frac{L_z}{\sin^2{\theta}} - aE
+ \frac{a}{\Delta}\left[  E(r^2+a^2) -L_za \right],\\
V_t &= a \left(L_z - aE\sin^2{\theta}\right)
+ \frac{r^2+a^2}{\Delta} \left[ E(r^2+a^2) - L_za \right].\label{potentials}
\end{aligned}
\end{equation}
To solve the equations of geodesic motions (\ref{geodesic}) numerically,  \textcolor{black}{it's convenient to} introduce two angular variables $\psi$ and $\chi$ in place of $r$ and $\theta$ to avoid possible singularities.
$\psi$ is defined as
\be
r=\frac{p}{1+e\cos\psi}.
\en
It's evident the  periastron and  apastron locate at $r_p=\frac{p}{1+e}$ and $r_a=\frac{p}{1-e}$. $\chi$ is defined as $z=\cos^2\theta=z_{-}\cos^2\chi$, where $z_{-}$ is given
by
\be
\beta(z_{+}-z)(z_{-}-z)
=   \beta z^{2}-z\left[Q+L_{z}^{2}+a^{2}(\mu^2-E^{2})\right] +Q,
\en
with $\beta=a^{2}(\mu^2-E^{2})$. The parameter $\chi$ ranges from 0 to $2\pi$. As $\chi$ varies from 0 to $2\pi$, $\theta$ goes from one turning point $\theta_{min}$ to the
other $\theta_{max}$ and back to $\theta_{min}$. We introduce an ``inclination angle'' to replace Carter constant $\cos\iota=L_z/\sqrt{L_z^2+Q}$. For $\iota=0$ or $\pi$, $Q$ equals to
zero, from the geodesic equation we learn the mass points remain on the equatorial plane. So $Q=0$ corresponds to the
equatorial orbits while $Q\neq0$ corresponds to inclined orbits. The radial potential can be expanded as
\begin{equation}
V_r=(\mu^2-E^{2})\,(r_{a}-r)\,(r-r_{p})\,(r-r_{3})\,(r-r_{4}) ,
\end{equation}
which facilitates writing the evolution equations for $\psi$ and $\chi$ as
\begin{equation}
\begin{aligned}
\frac{d \psi}{d t} &=\frac{\sqrt{\mu^2 - E^{2}}\,\left[(p -
r_{3}(1 - e)) - e(p + r_{3}(1 -
e)\cos\psi)\right]^{\frac{1}{2}}\,\left[(p - r_{4}(1 + e)) + e(p -
r_{4}(1 + e)\cos\psi)\right]^{\frac{1}{2}}}{\left[\gamma +
a^{2}\,E\,z(\chi)\right](1 - e^{2})},\\
\frac{d \chi}{d t} &=\frac{\sqrt{\beta\,\left[z_{+} -
z(\chi)\right]}}{\gamma + a^{2}\,E\,z(\chi)},
\end{aligned}
\end{equation}
where $\gamma = E\left[\left(r^2+ a^2\right)^{2}/\Delta - a^2 \right]-\frac{2 M r a L_{z}}{\Delta}$. Solving these equations allow us to know the positions of mass center of the binary at any time.

 To calculate GWs, we need to calculate the positions of the two compact objects in binary system relative to the mass center. The mass center of the binary system is free-falling around the SBH,
 thus locally the spacetime around the mass center can be treated as Minkowskian spacetime. Since for binary system the size of their internal orbit is much smaller than that of their external
 orbit, i.e., $\tilde{p}$ is much smaller than $p$, so the internal motions of binary are dominated by internal Newtonian gravity. For convenience, the two-body problem is reduced to one-body
 problem, the relative motion is described by the Lagrangian
\be \label{eq_iner_oribt}
L=\frac{1}{2}m_r\dot{\vec{\tilde{r}}}^2-V^{(i)}(\tilde{r}),
\en
where $m_r=\frac{m_1m_2}{m_1+m_2}$ is the reduced mass, and the potential $V^{(i)}(\tilde{r})=-\frac{m_1 m_2}{\tilde{r}}$ describes the internal gravity. \textcolor{black}{Due to symmetry, we know the binary stars' internal orbit is planar. For the bounded orbit, the radial coordinate can be parametrized as $\tilde{r}=\frac{\tilde{p}}{1+\tilde{e}\cos(\tilde{\theta})}$, where the sami-latus $\tilde{p}$ and eccentricity $\tilde{e}$ relate to energy and angular momentum through $\tilde{p}=\frac{L^2}{m_r\alpha}$ and $\tilde{e}=\sqrt{1+\frac{2 E L^2}{m_r\alpha}}$ . For the $\tilde{r}$ motion, we have
\be
\frac{d\tilde{r}}{dt}-\sqrt{\frac{2}{m_r}\left(E-V^{(i)}(\tilde{r})\right)-\frac{L^2}{m_r^2 \tilde{r}^2}}=0,
\en
from which it's easy to obtain the equation of motion of $\tilde{\theta}$. After solving out $\tilde{\theta}=\tilde{\theta}(t)$ and $\tilde{r}=\tilde{r}(t)$, the positions of the binary stars in the frame of mass center can then be given by 
\be
\vec{\tilde{r}}_1=\frac{m_2}{m_1+m_2} \vec{\tilde{r}},  \quad\quad\quad\quad\quad   \vec{\tilde{r}}_2=-\frac{m_1}{m_1+m_2}\vec{\tilde{r}}.
\en
Then the positions of the binary stars in the frame centered on the SBH is given by $\vec{r}_1=\vec{r}_C+\vec{\tilde{r}}_1$ and $\vec{r}_2=\vec{r}_C+\vec{\tilde{r}}_2$.}

Solving the aforementioned equations that determine the motions of mass center and the internal motions allow us to give the orbit of the binary. In Fig.\ref{orbit} we show the orbits of one of the
compact objects in the binary system.  From the lower three detail images in Fig.\ref{orbit} one sees that, due to internal motions the orbit have small fluctuations, and thus the orbits fluctuate
more frequently as mass increases. The reason roots in that if keeping other parameters fixed, the cycle of orbit becomes shorter as the mass of gravitational source increases. The fluctuations are
expected to give rise to higher-frequency GWs.

\subsection{the GW waveform}
In the weak-field situation, the metric can be decomposed as $g_{\mu\nu}=\eta_{\mu\nu}+h_{\mu\nu}$, where $\eta_{\mu\nu}$ is the flat metric and $h_{\mu\nu}$ is small
perturbation. By introducing the traceless tensor $\bar{h}^{\mu\nu}\equiv h^{\mu\nu}-(1/2)\eta^{\mu\nu}h$ with $h=\eta^{\mu\nu}h_{\mu\nu}$, then the Einstein equations can be
recast to the form
\begin{equation}
\Box \bar{h}^{\mu\nu}=-16\pi T^{\mu\nu}.
\end{equation}
Here $T^{\mu\nu}$ denotes the energy-momentum tensor of the source. In the slow
motion limit, the GW is given by the quadrupole-octupole formula
\begin{eqnarray}\label{quadoct}
\bar{h}^{jk} =
    \frac2{r} \left[ \ddot{I}^{jk} - 2 n_i \ddot{S}^{ijk} +
    n_i \dddot{M}^{ijk}\right]_{t'=t-r},\
\end{eqnarray}
with
\begin{equation}
\begin{aligned}
I^{jk}(t')& = \int x'^{j}x'^{k} T^{00}(t',{\bf x'}) d^3x',\\
S^{ijk}(t')
&= \int x'^{j}x'^{k} T^{0i}(t',{\bf x'}) d^3x' ,\\
M^{ijk}(t')
&= \int x'^i x'^j x'^k T^{00}(t',{\bf x'}) d^3x',\label{momenta}
\end{aligned}
\end{equation}
being the mass quadrupole moment, current quadrupole moment and mass octupole moment respectively, and $r^{2} = \bf{x}\cdot \bf{x}$, ${\bf n} = \bf{x} /r$, where $\bf{x}$ is
the location of the observer.

\begin{figure*}[h]
\begin{center}
\includegraphics[width=0.4 \textwidth]{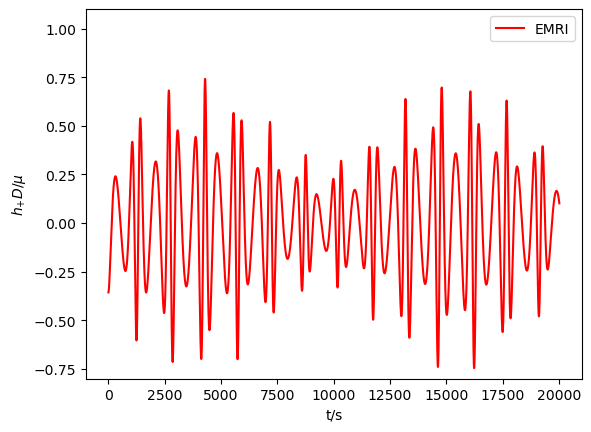}
\includegraphics[width=0.4 \textwidth]{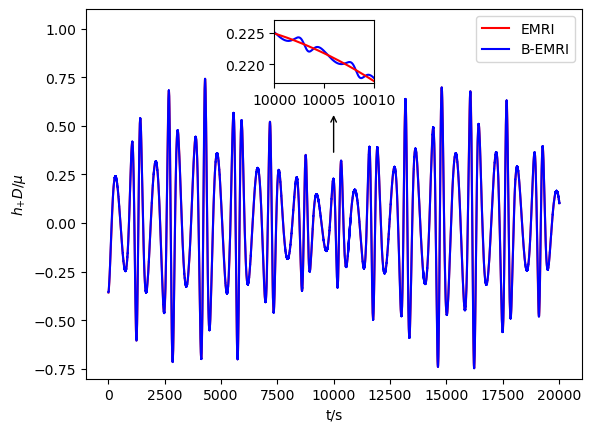}
\includegraphics[width=0.4 \textwidth]{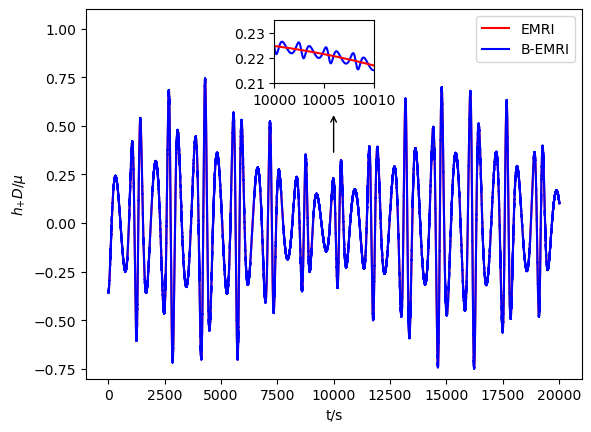}
\includegraphics[width=0.4 \textwidth]{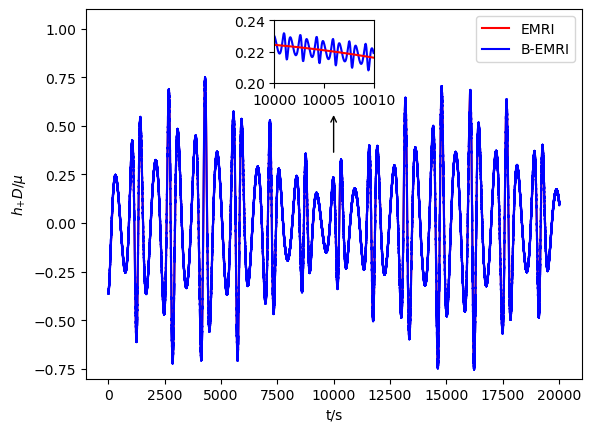}
\end{center}
\caption{$h_{+}$ waveforms of EMRIs and B-EMRIs with different masses. The parameters follow that given in Fig.\ref{orbit}, panel by panel.
}
\label{waveformsmass}
\end{figure*}

In the NK prescription, a flat-space trajectory which is
 ``equivalent'' to a geodesic in Kerr spacetime is reconstructed  by projecting the Boyer-Lindquist coordinates onto a fictitious spherical polar grid.
A particle moving along geodesic path in Kerr spacetime can be viewed as the particle moving along curved path in flat-space, so it's convenient for us to calculate GWs with Cartesian coordinates.
In Cartesian coordinates the positions of one of the compact objects is given by
\begin{equation}
\begin{aligned}
x_1&=r\sin\theta\cos\phi-\frac{m_2}{m_1+m_2}\frac{\tilde{p}}{1+\tilde{e}\cos\tilde{\beta}}\left(\sin(\tilde{\beta}+\gamma)\cos\phi_0+\cos(\tilde{\beta}
+\gamma)\cos\theta_0\sin\phi_0\right),\\
y_1&=r\sin\theta\sin\phi+\frac{m_2}{m_1+m_2}\frac{\tilde{p}}{1+\tilde{e}\cos\tilde{\beta}}\left(-\sin(\tilde{\beta}+\gamma)\sin\phi_0+\cos(\tilde{\beta}
+\gamma)\cos\theta_0\cos\phi_0\right),\\
z_1&=r\cos\theta+\frac{m_2}{m_1+m_2}\frac{\tilde{p}}{1+\tilde{e}\cos\tilde{\beta}}\cos(\tilde{\beta}+\gamma)\sin\theta_0,
\end{aligned}
\end{equation}
where $\gamma$ is the angle between the line connecting mass center and one of the compact objects and the line that lies in the plane of binary internal orbit and is perpendicular to the node line.
For the position of the other compact object one only has to replace $\frac{m_2}{m_1+m_2}$ with $-\frac{m_1}{m_1+m_2}$ in the above expression.

The expression (\ref{quadoct}) are valid for a general extended source in flat-space. For a single mass point with mass $m_1$  the energy-momentum tensor in flat spacetime is
given by
\begin{equation}
T^{\mu\nu} (t',{\bf x'})
=   \mu \int_{-\infty}^{\infty} \frac{\rm{d} x'^{\mu}_p}{\rm{d} \tau}
    \frac{\rm{d} x'^{\nu}_p}{\rm{d} \tau} \delta^{4}
    \left(x'-x'_{p}(\tau)\right) \;\; \rm{d} \tau
 =
    \mu \left(\frac{\rm{d} \tau}{\rm{d} t'_p}\right)^2 \frac{\rm{d} x'^{\mu}_p}{\rm{d} \tau}
    \frac{\rm{d} x'^{\nu}_p}{\rm{d} \tau}
    \delta^{3} \left( {\bf x}' - {\bf x}_{p}'(t') \right).
\end{equation}
For binary stars we know the energy-momentum tensor is the sum of two such terms with $\mu=m_1$ and $m_2$ respectively. The momenta (\ref{momenta}) for mass point take the simple form
\begin{equation}
\begin{aligned}
I^{jk} &= \mu  x'^j_p x'^{k}_p\;,
\\
S^{ijk} &= v^i\ I^{jk}\;,
\\
M^{ijk} &= x'^i_p\ I^{jk}\;.\label{momenta2}
\end{aligned}
\end{equation}
Substituting (\ref{momenta2}) into (\ref{quadoct}) enables to work out the GWs for a mass point. For binary system the GW is the superposition of that generated by two compact objects.

\begin{figure*}
\begin{center}
\includegraphics[width=0.32 \textwidth]{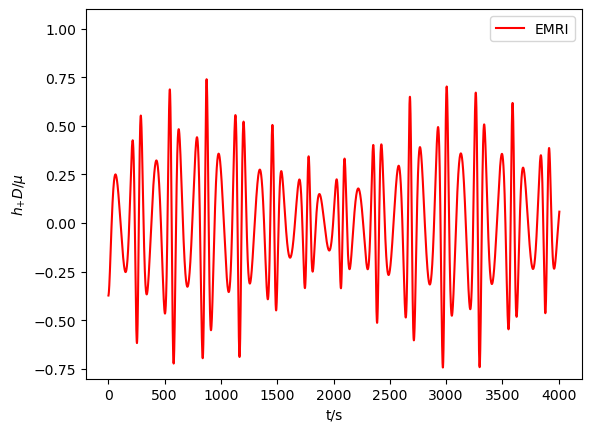}
\includegraphics[width=0.32 \textwidth]{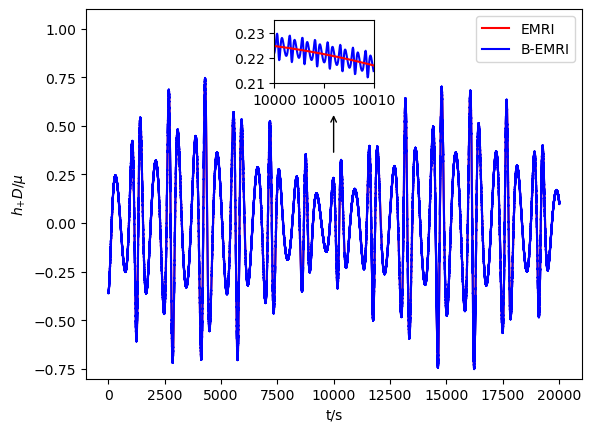}
\includegraphics[width=0.32 \textwidth]{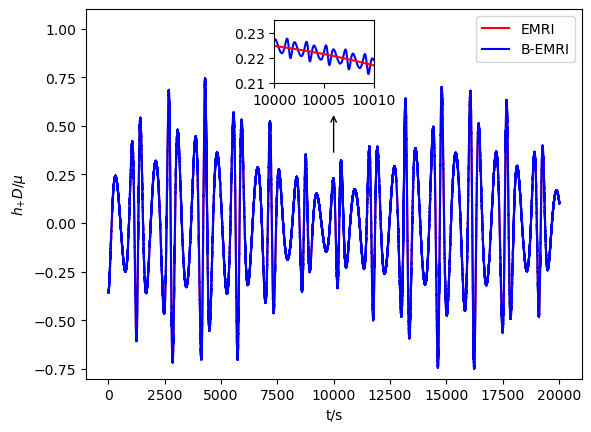}
\includegraphics[width=0.32 \textwidth]{Binary-p10-p0005-810.png}
\includegraphics[width=0.32 \textwidth]{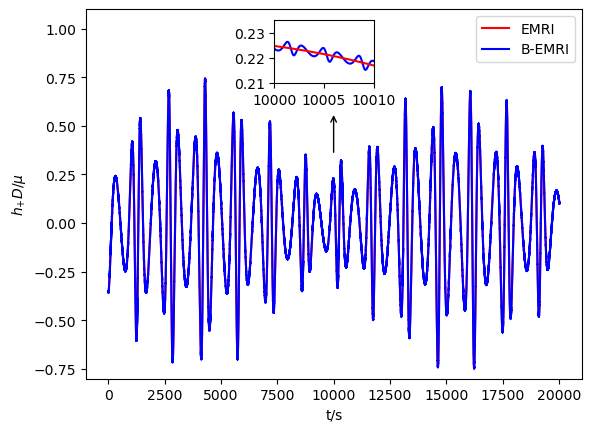}
\includegraphics[width=0.32 \textwidth]{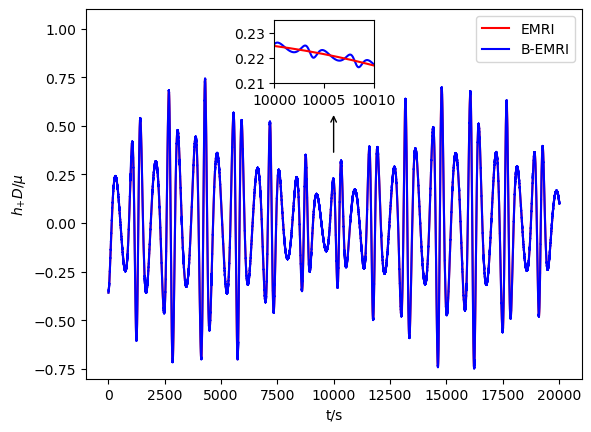}
\end{center}
\caption{Comparisons between the $h_{+}$ waveforms of EMRIs and B-EMRIs
    with distinct semi-latus rectum $\tilde{p}$.
  The  binary masses are fixed to be $m_1=8M_{\odot}$ and $m_2=10 M_{\odot}$, and mass of central SBH is $10^6M_{\odot}$. The external orbit parameters are set to be $p(0)=10M, e(0)=0.3,
  \iota(0)=7\pi/9,
 a=0.9M$, the internal orbit parameters of the binary stars are set to be $\tilde{e}=0.2, \theta_0=\frac{\pi}{3}$ and $\tilde{p}=0.003M, 0.004M, 0.005M, 0.006M, 0.007M$ respectively.
 }\label{waveformsp}
\end{figure*}

In the standard transverse-traceless (TT) gauge, the TT projection of $\bar{h}^{jk}$ gives rise to
\begin{equation}\label{htt}
h^{jk}_{TT} = \frac1{2}\left(
       \begin{array}{ccc}
       0 & 0 & 0\\
       0 & h^{\Theta\Theta}-h^{\Phi\Phi}& 2h^{\Theta\Phi}\\
       0 & 2h^{\Theta\Phi} & h^{\Phi\Phi}-h^{\Theta\Theta}
        \end{array}
 \right),
\end{equation}
with
\begin{equation}
\begin{aligned}
h^{\Theta\Theta}
&= \cos^2{\Theta} \left[ h^{xx} \cos^2{\Phi} + h^{xy}\sin{2\Phi} +
    h^{yy}\sin^2{\Phi} \right] + h^{zz}\sin^2{\Theta} -
    \sin{2\Theta}\left[ h^{xz}\cos{\Phi} + h^{yz}\sin{\Phi}\right],\\
h^{\Phi\Theta}
&= \cos{\Theta} \left[ -\frac1{2} h^{xx}\sin{2\Phi} +
    h^{xy}\cos{2\Phi} +\frac1{2}h^{yy}\sin{2\Phi} \right] +
    \sin{\Theta}\left[ h^{xz}\sin{\Phi} - h^{yz}\cos{\Phi} \right],\\
h^{\Phi\Phi}
&= h^{xx}\sin^2{\Phi} - h^{xy}\sin{2\Phi} + h^{yy}\cos^2{\Phi}.
\end{aligned}
\end{equation}
From (\ref{htt}) it's easy to see the two polarizations states read
$h_{+}=h^{\Theta\Theta}-h^{\Phi\Phi}$ and $h_{\times}=2h^{\Theta\Phi}$.

Due to the GW emissions, the compact objects will lose energy and angular momentum, although the amount of energy and angular momentum carried away by GWs is very small, the trajectory will alter
accordingly, for a more precise study we consider backreaction of energy loss to the orbit. We study generic inclined-eccentric orbits for the mass center of binary stars. As we will see, since the amplitude of GW generated by internal motions of binary stars is small compared to that generated by external motions,
we  consider the radiation reaction caused by the external motion of binary stars around the SBH while that caused by the internal motion is neglected. We consider the flux
of generic orbits to 2nd order post-Newtonian approximation\cite{Gair:2005ih}
\begin{equation}
\begin{aligned}
(\dot{E})_{\rm 2PN} &= -\frac{32}{5} \frac{\mu^2}{M^2} \left(\frac{M}{p}\right)^5
(1-e^2)^{3/2}\left [ g_1(e) -q\left(\frac{M}{p}\right)^{3/2} g_2(e)
\cos\iota -\left(\frac{M}{p}\right) g_3(e) +
\pi\left(\frac{M}{p}\right)^{3/2} g_4(e)  \right.  \\
& \left. -\left(\frac{M}{p}\right)^2 g_5(e)
+  q^2\left(\frac{M}{p}\right)^2 g_6(e)
 -\frac{527}{96} q^2 \left( \frac{M}{p}\right)^2 \sin^2\iota
   \right ] ,\
  \\
(\dot{L}_z)_{\rm 2PN} &= -\frac{32}{5} \frac{\mu^2}{M} \left(\frac{M}{p}\right)^{7/2}
(1-e^2)^{3/2}
\left[ g_9(e)\cos\iota  + q\left(\frac{M}{p}\right)^{3/2} \{g_{10}^{a}(e)
-\cos^2\iota g_{10}^{b}(e) \}
-\left(\frac{M}{p}\right) g_{11}(e) \cos\iota
\right.\\
&\left. + \pi\left(\frac{M}{p}\right)^{3/2} g_{12}(e)\cos\iota
-\left(\frac{M}{p}\right)^2 g_{13}(e) \cos\iota
+ q^2\left(\frac{M}{p}\right)^2  \cos\iota \left(g_{14}(e) - \frac{45}{8}\,\sin^2\iota\right)   \right],\\
(\dot{Q})_{\rm 2PN}&= -\frac{64}{5} \frac{\mu^2}{M} \left(\frac{M}{p}\right)^{7/2} \,\sqrt{Q}\,\sin\iota\,
(1-e^2)^{3/2}\,\left[ g_9(e) - q\left(\frac{M}{p}\right)^{3/2} \cos\iota g_{10}^{b}(e) -\left(\frac{M}{p}\right) g_{11}(e)
\right. \\
& \left. + \pi\left(\frac{M}{p}\right)^{3/2} g_{12}(e)
-\left(\frac{M}{p}\right)^2 g_{13}(e)
+ q^2\left(\frac{M}{p}\right)^2 \,\left(g_{14}(e) - \frac{45}{8}\,\sin^2\iota\right)  \right],\label{ELQdot}
\end{aligned}
\end{equation}
where $q=a/M$.
The coefficients $g_1(e)\sim g_{14}(e)$ are deferred to the Appendix.
This expression of fluxes are the results of combining Tagoshi's\cite{Tagoshi:1995sh}, Shibata's \cite{Shibata:1994jx} and Ryan's fluxes \cite{Ryan:1995zm,Ryan:1995xi}. Numerical analysis indicates
that these results
ameliorate the unphysical inspiral properties for nearly circular and polar orbits thus produce more physical reasonable results\cite{Gair:2005ih}. The evolutions of conserved quantities (\ref{ELQdot}) allow us to give the evolution of eccentricity and semi-latus rectum as
\begin{equation}
\begin{aligned}
\dot{e}&=\frac{\partial e}{\partial E}\dot{E}+\frac{\partial e}{\partial Lz}\dot{Lz}+\frac{\partial e}{\partial Q}\dot{Q},\\
\dot{p}&=\frac{\partial p}{\partial E}\dot{E}+\frac{\partial p}{\partial Lz}\dot{Lz}+\frac{\partial p}{\partial Q}\dot{Q},\\
\end{aligned}
\end{equation}
after some algebraic manipulation we obtain
\begin{equation}
\begin{aligned}
\dot{e}=&\frac{1-e^2}{2p}\left((1+e)\frac{N_1(r_p)}{N(r_p)}-(1-e)\frac{N_1(r_a)}{N(r_a)}\right)\dot{E}+\frac{1-e^2}{2p}\left((1+e)\frac{N_2(r_p)}{N(r_p)}-(1-e)\frac{N_2(r_a)}{N(r_a)}\right)\dot{L}\\
&+\frac{1-e^2}{2p}\left((1+e)\frac{N_3(r_p)}{N(r_p)}-(1-e)\frac{N_3(r_a)}{N(r_a)}\right)\dot{\iota},\\
\dot{p}=&\left(-\frac{(1+e)^2}{2}\frac{N_1(r_p)}{N(r_p)}-\frac{(1-e)^2}{2}\frac{N_1(r_a)}{N(r_a)}\right)\dot{E}+\left(-\frac{(1+e)^2}{2}\frac{N_2(r_p)}{N(r_p)}-\frac{(1-e)^2}{2}\frac{N_2(r_a)}{N(r_a)}\right)\dot{L}\\
&+\left(-\frac{(1+e)^2}{2}\frac{N_3(r_p)}{N(r_p)}-\frac{(1-e)^2}{2}\frac{N_3(r_a)}{N(r_a)}\right)\dot{\iota},
\end{aligned}
\end{equation}
where
\begin{equation}
\begin{aligned}
N_1(r)&=E r^4+a^2 E r^2-2a M (L_z-a E) r,\\
N_2(r)&=-L_z\sec^2\iota\;r^2+2M (L_z\sec^2\iota-a E)r-a^2L_z\tan^2\iota,\\
N_3(r)&=L_z^2\tan\iota\sec^2\iota(r(2M-r)-a^2), \\
N(r)&=-2 (\mu^2-E^2) r^3+3\mu^2 M r^2-(a^2 (\mu^2-E^2)+L_z^2\sec^2\iota) r+L_z^2M\tan^2\iota+(L_z-a E)^2M.
\end{aligned}
\end{equation}

\begin{figure*}
\begin{center}
\includegraphics[width=0.32 \textwidth]{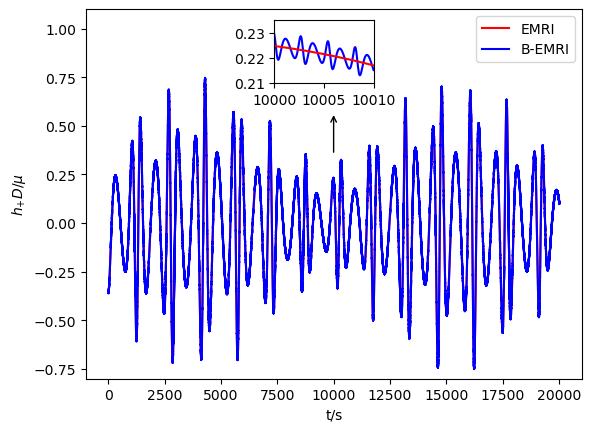}
\includegraphics[width=0.32 \textwidth]{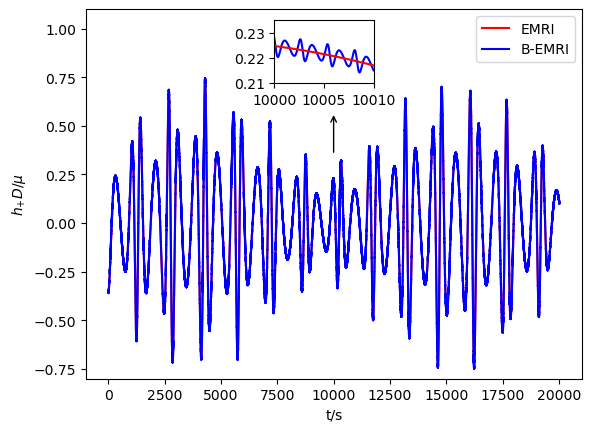}
\includegraphics[width=0.32 \textwidth]{Binary-p10-p0005-810.png}
\includegraphics[width=0.32 \textwidth]{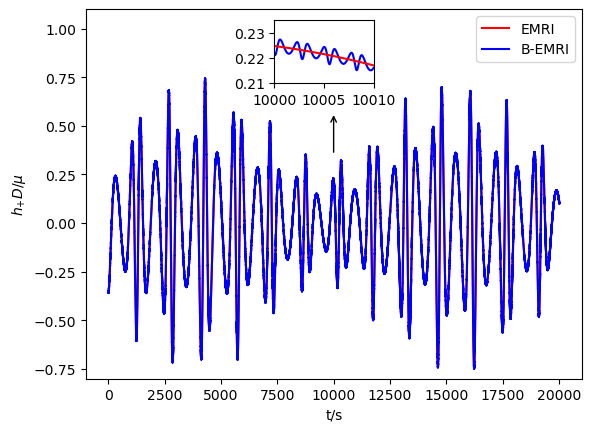}
\includegraphics[width=0.32\textwidth]{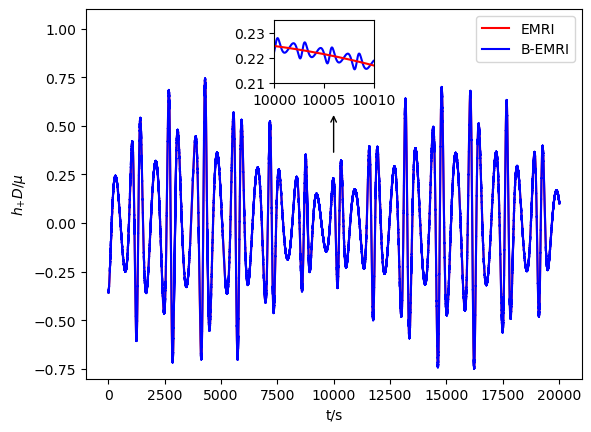}
\includegraphics[width=0.32 \textwidth]{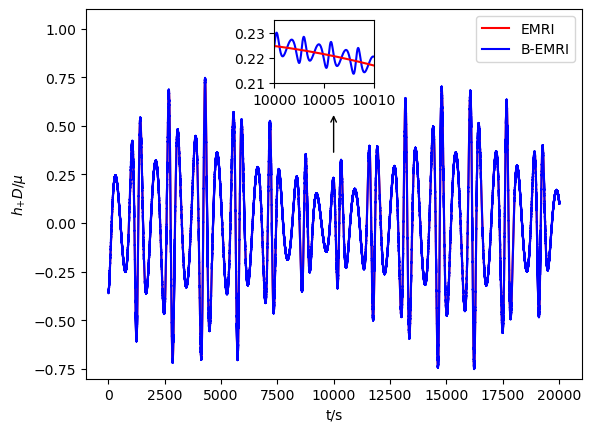}
\end{center}
\caption{Comparisons of $h_{+}$ waveforms of EMRIs and B-EMRIs by adjusting $\theta_0$.  The external orbit parameters are set to be $p(0)=10M, e(0)=0.3, \iota(0)=7\pi/9,
 a=0.9M$, the internal orbit parameters are set to be $\tilde{p}=0.005M, \tilde{e}=0.2$, and $\theta_0=0, \frac{\pi}{6}, \frac{\pi}{3}, \frac{\pi}{2}, \frac{3\pi}{4}, \pi$ respectivly.
}\label{waveformstheta}
\end{figure*}

Using the relation between inclination angle and Carter constant $Q=L_z^2\tan^2\iota$, one has
\be\label{iotadot}
\dot{\iota}=\frac{1}{2 L_z^2\tan(\iota)\sec^2(\iota)}\dot{Q}-\frac{\sin(\iota)\cos(\iota)}{L_z}\dot{L_z}.
\en

\begin{figure*}[h]
\begin{center}
\includegraphics[width=0.4 \textwidth]{Emris-single.png}
\includegraphics[width=0.4 \textwidth]{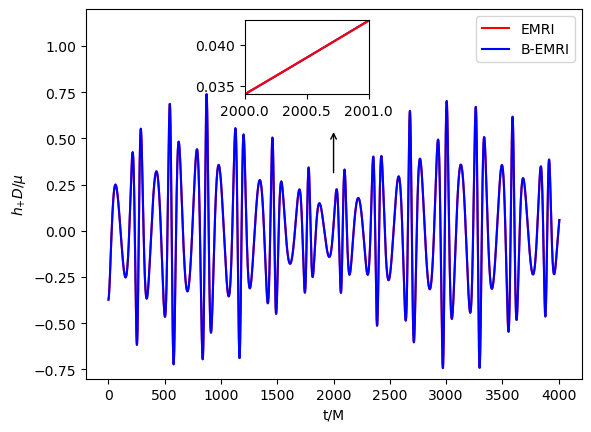}
\includegraphics[width=0.4 \textwidth]{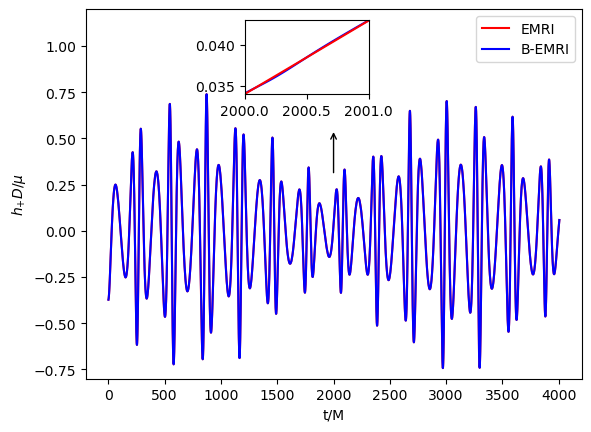}
\includegraphics[width=0.4 \textwidth]{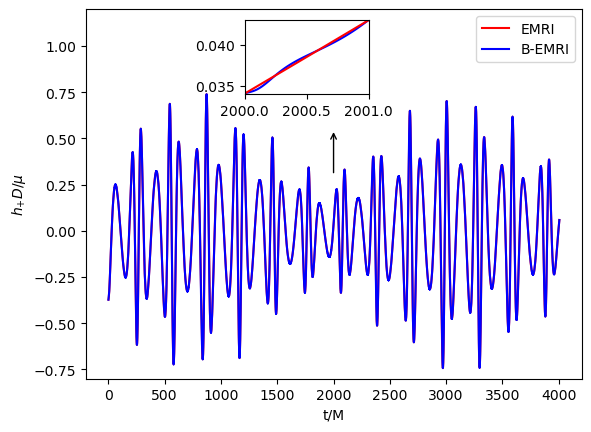}
\end{center}
\caption{$h_{+}$ waveforms of EMRIs and B-EMRIs with central BH of mass $10^8M_{\odot}$. The masses of the binary are  $8M_{\odot}$ and $10M_{\odot}$ (left panel),  $30M_{\odot}$ and  $30M_{\odot}$
(middle panel), and $80M_{\odot}$ and $100M_{\odot}$ (right panel). The external orbit parameters are set to be $p(0)=10M, e(0)=0.3, \iota(0)=7\pi/9$, the internal motion parameters are set to be
$\tilde{p}=0.004M$ and $\tilde{e}=0.2$.
}\label{waveformsyimass}
\end{figure*}

Solving all the evolution equations above with the NK numerical method enables us to calculate the GWs. In Figs.\ref{waveformsmass}-\ref{waveformsyimass}, we exhibit the $h_{+}$ waveforms of EMRIs and B-EMRIs
for different parameter selections. The orientation of the internal elliptic orbit plane is determined by the angles  $\theta_0$ and $\phi_0$, $\phi_0$ is fixed to be $\frac{\pi}{6}$. The distance
between the observer and the source is supposed to be $r=D$. From the figure we see that, the waveforms of B-EMRIs have the identical profile as that of EMRIs, but the waveforms of B-EMRIs have
oscillations which are caused by the internal motion of the binary. We will analyze the waveforms in detail in the next section.

\textcolor{black}{We should analyze whether our system is stable or whether the binary stars will be disrupted tidally by the central SBH. If the binary star system approaches the SBH closer than $r_t$, it will be disrupted, $r_t$ is given by
\be
\frac{M}{r_t^3}\sim \frac{m_1+m_2}{\tilde{a}^3}, 
\en
where $\tilde{a}$ is the semi-major axis of the binary internal orbit. To estimate the tidal disruption radius, we take $\tilde{a}=0.005M$, $M=10^6 M_{\odot}$ and $m_1+m_2=18 M_{\odot}$ according to the paper, with these parameter selections we obtain $r_t=0.19M$, which is much less than the semi-latus rectum of the external orbit $p=10M$, so the binary system is stable against the tidal force of the SBH.}

Before proceeding, let's do some analytic calculations to compare with our numerical results. To isolate the dominant physical mechanisms, we omitted the higher-order back reaction terms. We consider a typical case,
for which the binary's external orbit is a circular one on the equatorial plane ($x-y$ plane) of the SBH, the masses of the binary are equal, their internal orbit is circular and is on the
equatorial plane as well. The mass of the SBH is set to be $10^6M_{\odot}$, the mass of the binary both are set to be $8M_{\odot}$, the radius of external orbit is $R_0=10 M$, and the radius of
internal orbit is $d=0.005 M$.

The positions of the binary are given by
\begin{equation}
\begin{aligned}
x_1&=R_0 \cos(\omega_0t)+d\cos(\omega t),\\
y_1&=R_0 \sin(\omega_0t)+d\sin(\omega t),\\
x_2&=R_0 \cos(\omega_0t)-d\cos(\omega t),\\
y_2&=R_0 \sin(\omega_0t)-d\sin(\omega t),
\end{aligned}
\end{equation}
where $\omega_0=V_{\phi}/V_t$ is the circular frequency of the binary's external orbit, and $\omega$ is the circular frequency of the internal orbit.

Fundamentally, we consider the contribution of mass quadrupole moment as the first step,
\begin{equation}
\bar{h}^{jk}=\frac{2}{r}\ddot{I}^{jk},
\end{equation}
straightforward calculations give rise to
\begin{equation}
h_{+}=-\frac{10\mu}{r}\left[R_0^2\omega_0^2\cos(2\omega_0 t)+d^2\omega^2\cos(2\omega t)\right],
\end{equation}
where the observation point's latitude and azimuth are set to be $\Theta=\frac{\pi}{3}, \Phi=0$ as above.

\begin{figure*}
\begin{center}
\includegraphics[width=0.45 \textwidth]{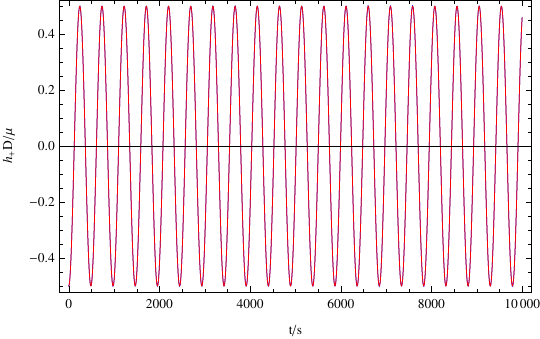}
\includegraphics[width=0.45 \textwidth]{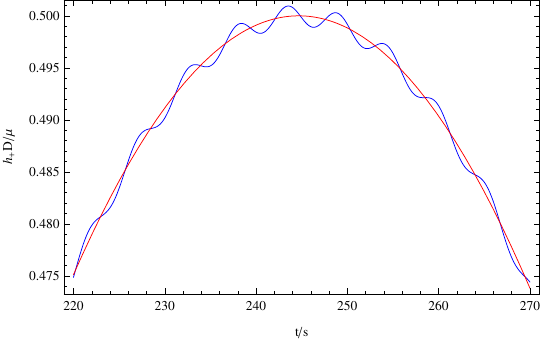}
\end{center}
\caption{$h_{+}$ waveforms of EMRIs (red) and B-EMRIs (blue) calculated with mass quadrupole moment.} \label{hpI}
\end{figure*}

For the EMRIs with single small compact object, the position of the small body is given by
\begin{equation}
x=R_0\cos(\omega_0t),\quad\quad y=R_0\sin(\omega_0t).
\end{equation}
Taking into account the contribution of mass quadrupole moment, straightforward calculations give rise to
\begin{equation}
h_{+}=-\frac{10\mu}{r} R_0^2\omega_0^2\cos(2\omega_0 t).
\end{equation}
The waveforms of EMRIs and B-EMRIs by  taking into account the contribution of mass quadrupole moment are shown in Fig.\ref{hpI}.

\begin{figure*}
\begin{center}
\includegraphics[width=0.45 \textwidth]{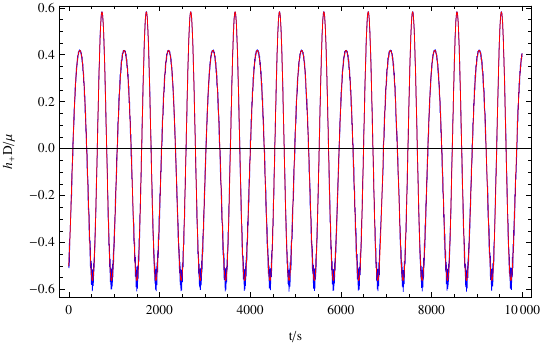}
\includegraphics[width=0.45 \textwidth]{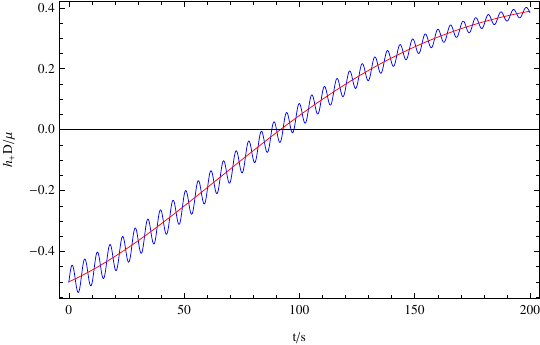}
\end{center}
\caption{$h_{+}$ waveforms of EMRIs (red) and B-EMRIs (blue) calculated with mass quadrupole moment, current quadrupole moment and mass mass octupole moment.} \label{hp}
\end{figure*}

If incorporating the contributions of current quadrupole moment and mass mass octupole moment as in Eq.(\ref{quadoct}),
the GW of B-EMRIs is given by
\begin{equation}
\begin{aligned}
h_{+}&=\frac{\mu}{8r}\left\{-80 d^2 \omega^2 \cos(2 \omega t )   -80 R_0^2 \omega_0^2 \cos(2 \omega_0 t)
+ \sqrt{3}R_0 \left[d^2 (-2 \omega + \omega_0)^2 \right.\right.\\
&\left.\left.\cdot(10 \omega + 11 \omega_0) \sin((2 \omega - \omega_0)t)
 + 2 \omega_0^2 \left(33 R_0^2 \omega_0 + d^2 (16 \omega + 5 \omega_0)\right.\right.\right.\\
  &\left.\left.\left.+ 45 R_0^2 \omega_0 \cos(2 \omega_0 t)\right) \sin(\omega_0 t) + 5 d^2 (2 \omega + \omega_0)^3 \sin((2 \omega + \omega_0)t)\right]\right\},
\end{aligned}
\end{equation}
and that of EMRIs is given by
\begin{equation}
h_{+}=\frac{\mu}{8r}R_0^2\omega_0^2\left[-80\cos(2\omega_0t)+3\sqrt{3}R_0\omega_0(7\sin(\omega_0t)+15\sin(3\omega_0t))\right]
\end{equation}
The quadrupole-octupole waveforms of EMRIs and B-EMRIs are shown in Fig.\ref{hp}.

\begin{figure*}
\begin{center}
\includegraphics[width=0.4 \textwidth]{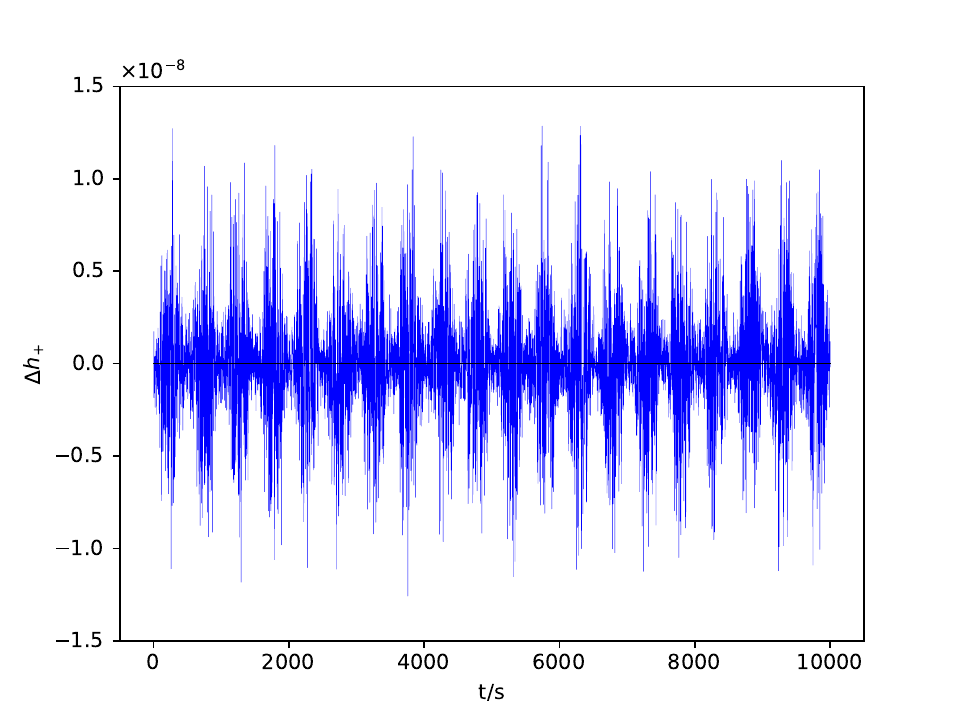}
\includegraphics[width=0.4 \textwidth]{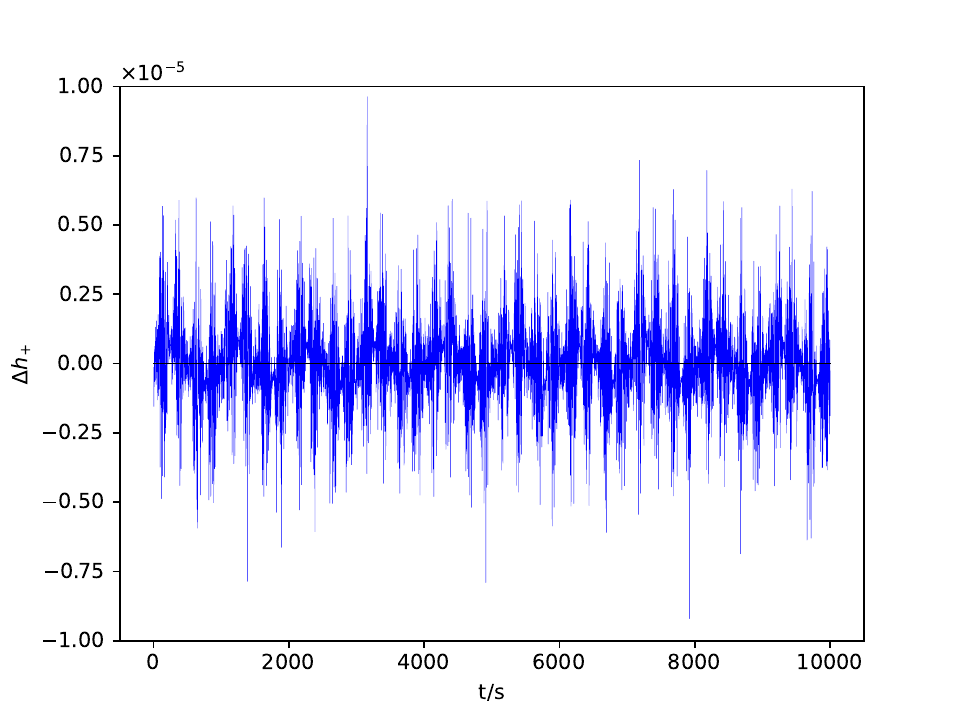}
\end{center}
\caption{This figure illustrates the differences between the analytical and numerical waveforms of the mass quadrupole moment and that including current quadrupole moment and mass octupole moment.} \label{diff1}
\end{figure*}

To further investigate the reliability of numerical calculations in the text, we carefully
compared the numerical results with the aforementioned analytical solutions in Fig.\ref{diff1}. Our conclusion demonstrates that the numerical results and analytical solutions
mutually corroborate and exhibit a high degree of agreement, eff ectively capturing the
key physical factors. When considering higher-order terms, the consistency between
numerical and analytical results slightly decreases due to the introduction of higher-
order derivatives, yet still fully meets the observational accuracy requirements.

\subsection{Gravito-electromagnetic force}
The internal orbit is calculated  by  Newtonian  dynamics in Eq. \ref{eq_iner_oribt}.  It is interesting and important to consider the higher order corrections e.g. the Gravito-electromagnetic force
effect \cite{Chen:2022tdj}.   Unlike single free-falling compact object which always locates at the origin of the free-fall frame (FFF), every compact objects of the binary system will not locate at
the origin of FFF due to the internal interaction of binary.
According to equivalence principle only at the origin of FFF the spacetime is flat, if there is departure from the origin gravity will play a role. So for the binary we take into account the
gravitational effects caused by the departure from the origin of FFF. For simplicity, we consider a circular orbit on the equatorial plane in the following. The natural choice of FFF is  Riemann
normal coordinates
\be
ds^2=-\left(1+R_{0i0j}x^ix^j\right)c^2d\tau^2-\frac{4}{3}R_{0jik}x^jx^k d\tau dx^i+\left(\delta_{ij}-\frac{1}{3}R_{ikjl}x^kx^l\right)dx^idx^j.
\en
If $R_{ikjl}$, which is small compared to $R_{0i0j}$ and $R_{0jik}$, is neglected, the metric above can be rewritten as
\be
ds^2=-\left(1-2\frac{\Phi}{c^2}\right)c^2d\tau^2-\frac{4}{c}(\vec{A}\cdot d\vec{x}) cd\tau +\delta_{ij}dx^idx^j,
\en
where
\begin{equation}
\begin{aligned}
\Phi&=-\frac{1}{2}R_{0i0j}x^ix^j,\\
A_i&=\frac{1}{3}R_{0jik}x^jx^k,
\end{aligned}
\end{equation}
are the scalar and vector potentials of GEM fields. Analogous to the electromagnetic fields, the GEM force can be introduced
\be
\vec{F}=-m\vec{E}-2m \frac{\vec{v}}{c}\times \vec{B},
\en
where
\begin{equation}
\begin{aligned}
E_i&=R_{0i0j}x^j,\\
B_i&=-\frac{1}{2}\epsilon_{ijk}R_{0l}^{\;\;\; jk}x^l.
\end{aligned}
\end{equation}
In FFF the equations of motion of the compact object in binary system can be wriiten as
\be\label{GEMeom}
m_i\frac{d^2\vec{x}_i}{d\tau^2}=-m_i m_j \frac{\vec{x}_i-\vec{x}_j}{|\vec{x}_i-\vec{x}_j|^3}+\vec{F}_i
\en
where $i=1,2$ label the compact objects which constitute the binary.

\begin{figure*}
\begin{center}
\includegraphics[width=0.4 \textwidth]{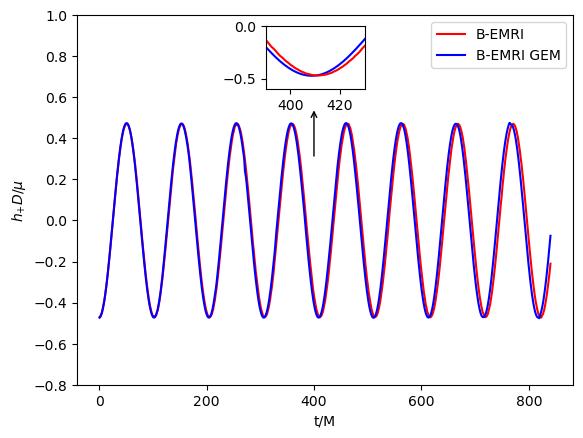}
\includegraphics[width=0.4 \textwidth]{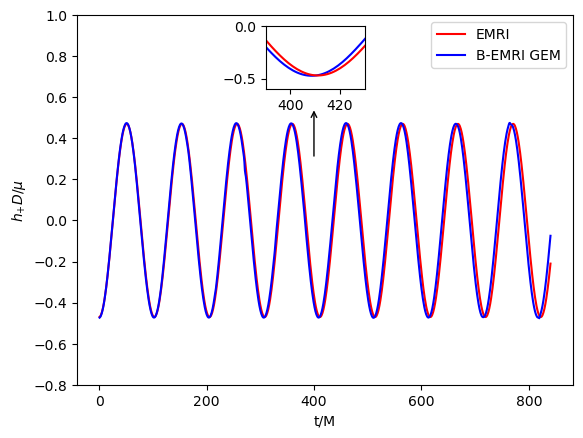}
\end{center}
\caption{GWs generated by B-EMRIs on the equatorial plane with and without taking into account gravito-electromagnetic force. The mass of central SBH is set to be $10^8M_{\odot}$, and the masses of
the binary are set to be $8M_{\odot}$ and $10M_{\odot}$. The external orbit parameters are set to be $p(0)=10M, e(0)=0.3, a=0.9M$, the semi-major axis and eccentricity of internal orbit are set to
be $\tilde{a}=10^5(m_1+m_2), \tilde{e}=0.2$. The frequency of GWs is slightly increased when GEM force is in consideration. The physical interpretation is that the effect of general
relativity tends to strengthen gravity compared to the Newtonian gravity.
}\label{Binary-GEM}
\end{figure*}

\begin{figure*}
\begin{center}
\includegraphics[width=0.3 \textwidth]{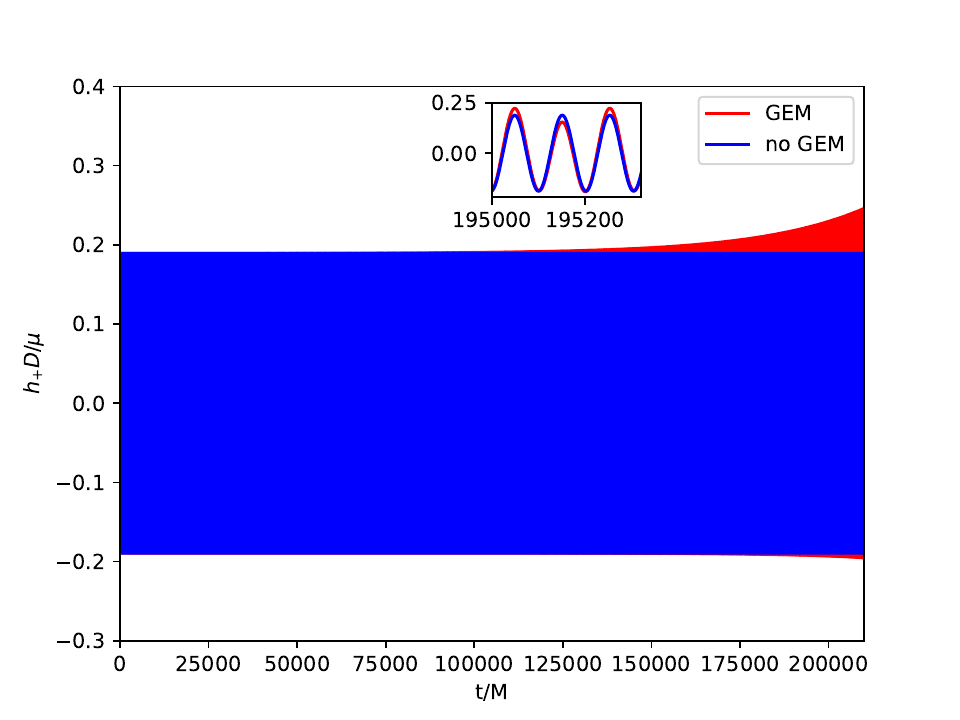}
\includegraphics[width=0.3 \textwidth]{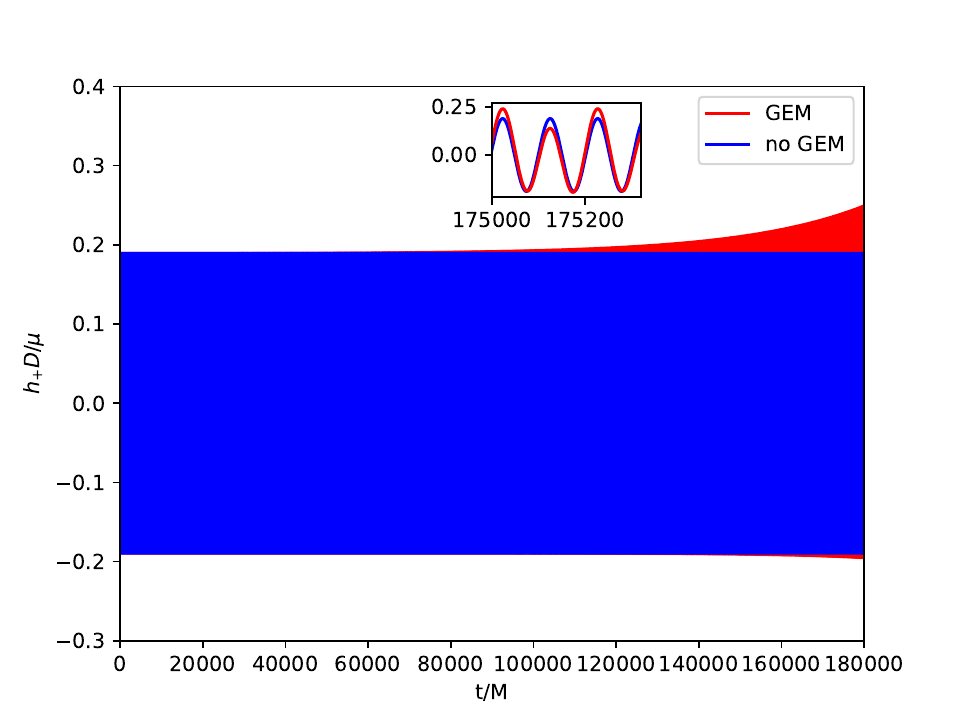}
\includegraphics[width=0.3 \textwidth]{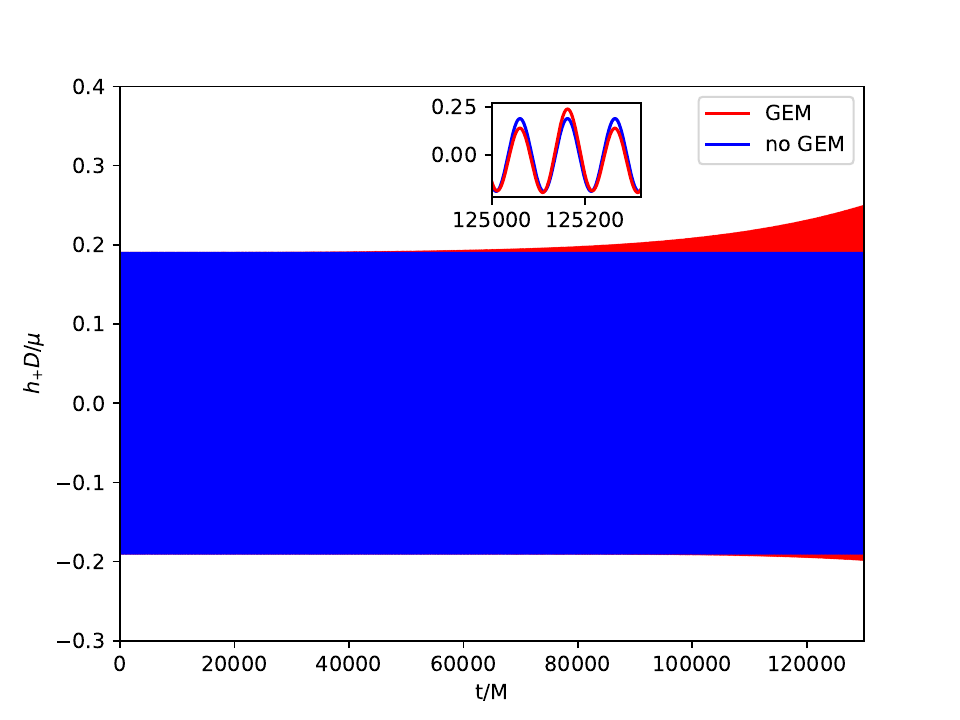}
\end{center}
\caption{\textcolor{black}{GWs generated by B-EMRIs on the equatorial plane with and without taking into account gravito-electromagnetic force. The mass of central SBH is set to be $4\times10^6M_{\odot}$, and the masses of
the binary stars are set to be $m_1=8M_{\odot}$ and $m_2=10M_{\odot}$. The external circular orbit parameters are set to be $r=10M, a=0.9M$, semi-major axis of the internal orbit is set to
be $\tilde{a}=10^5(m_1+m_2)$, eccentricities of the internal orbits are $\tilde{e}=0.2, 0.5, 0.8$ respectively for the three panels.}
}\label{Binary-GEM-106sun}
\end{figure*}

Although the equations of motion can be written in a simple form in FFF, it is not straightforward to calculate GEM force in FFF since Riemann tensor is conventionally derived in local non-rotating
frame (LNRF). The transformations between the coordinates of LNRF ($X^t, X^r, X^\theta, X^\phi$) and BL coordinates are given by
\begin{equation}
\begin{aligned}
dX^t&=(\Sigma\Delta/A)^{1/2}dt,\\
dX^r&=(\Sigma/\Delta)^{1/2}dr,\\
dX^\theta &=\Sigma^{1/2}d\theta,\\
dX^\phi &=-\frac{2Mar\sin\theta}{(\Sigma A)^{1/2}}+(A/\Sigma)^{1/2}\sin\theta d\phi,
\end{aligned}
\end{equation}
where $A=(r^2+a^2)^2-a^2\Delta\sin^2\theta$. It's shown Riemann tensor in LNRF takes very simple form\cite{Bardeen:1972fi,Chandrasekhar}. For circular orbit on the equatorial plane, the free-fall
observer is moving in the $X^\phi$ direction with the speed\cite{Bardeen:1972fi}
\be
u=\frac{\pm M^{1/2}(r\mp 2a M^{1/2} r^{1/2}+a^2)}{\Delta^{1/2}(r^{3/2}\pm aM^{1/2})}
\en
where the upper signs refer to prograde orbits and the lower ones refer to retrograde orbits. The LNRF and the local inertial frame (LIF) are connected by a Lorentz transformation
\be
\Lambda^\mu_{\mu'}=\left(
    \begin{array}{cccc}
        \gamma & 0 & 0 & -\gamma\beta\\
       0  & 1 & 0 &0 \\
       0  & 0 & 1 & 0\\
        -\gamma \beta & 0 & 0 & \gamma\\
        \end{array}
   \right), \label{lorentz}
\en
where $\beta=u/c, \gamma=\sqrt{1-\beta^2}$. Through the Lorentz transformation (\ref{lorentz}) Riemann tensor in LNRF can be transformed to the ones in LIF
$R_{\mu'\nu'\rho'\sigma'}=\Lambda^\mu_{\mu'}\Lambda^\nu_{\nu'}\Lambda^\rho_{\rho'}\Lambda^\sigma_{\sigma'}R_{\mu\nu\rho\sigma}$. If a free gyro placed at the origin of FFF, its spin axis is fixed in
FFF, but in general the gyro will precess relative to the LNRF and LIF, so there exist rotation between LIF and FFF, the angular velocity is $\omega=-\sqrt{M/r^3}$\cite{RindlerPerlick}. The tensors
in LIF and the that in FFF are connected through a rotation transformation. The components of GEM force in FFF can be given in terms of that in LIF as
\begin{equation}
\begin{aligned}
F_x&=F_{r'}\cos(\omega\tau)+F_{\phi'}\sin(\omega\tau),\\
F_y&=F_{\theta'},\\
F_z&=-F_{r'}\sin(\omega\tau)+F_{\phi'}\cos(\omega\tau).\label{forceFFF}
\end{aligned}
\end{equation}
Substituting the GEM force (\ref{forceFFF}) into the equations of motion (\ref{GEMeom}) and solving the equations allows  to obtain the positions of compact objects in FFF. Considering the orbit of
origin of FFF is geodesic in external SBH spacetime, it's straightforward to obtain the positions of compact objects of the binary in SBH spacetime, based on this one is able to calculate the GWs
with the NK method. In Figs.\ref{Binary-GEM} and \ref{Binary-GEM-106sun} we present the waveforms of GWs with and without taking into account GEM force.

\section{Analysis of the waveforms\label{section3}}
In this section, we compare the waveforms of B-EMRIs with that of EMRIs in detail. Furthermore, we investigate the modification of the waveform by considering  the Gravito-electromagnetic (GEM)
force effect and show that the dominated effect of GEM is a surplus phase shift. Fig.\ref{waveformsmass} displays the $h_{+}$ waveforms of EMRIs and B-EMRIs with different masses.  One sees that the
profile of GWs of EMRIs and B-EMRIs are similar, but the details are distinct. For B-EMRIs, the gravitational waveforms possess small fluctuations. This is caused by the internal motions of the
binary. We see, as mass increases, both the amplitudes and the frequency of the fluctuations increase.
Fig.\ref{waveformsp} displays the the $h_{+}$ waveforms of EMRIs and B-EMRIs with different semi-latus rectum $\tilde{p}$ of the internal orbit. The figure shows that the smaller $\tilde{p}$ is, the
larger the amplitude and the frequency of the fluctuation are, and vice versa. Since smaller $\tilde{p}$ corresponds to small cycle or higher frequency, amplitude of the fluctuation increases as
$\tilde{p}$ decreases is expected.

\begin{figure*}[h]
\begin{center}
\includegraphics[width=0.4 \textwidth]{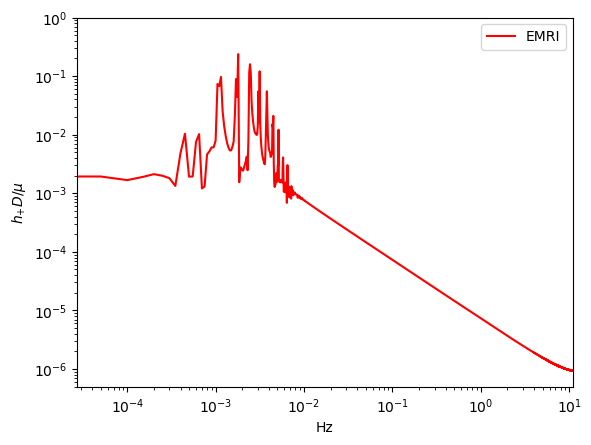}
\includegraphics[width=0.4 \textwidth]{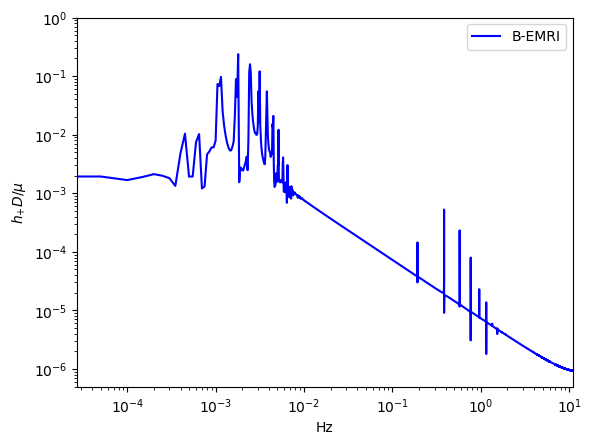}
\includegraphics[width=0.4 \textwidth]{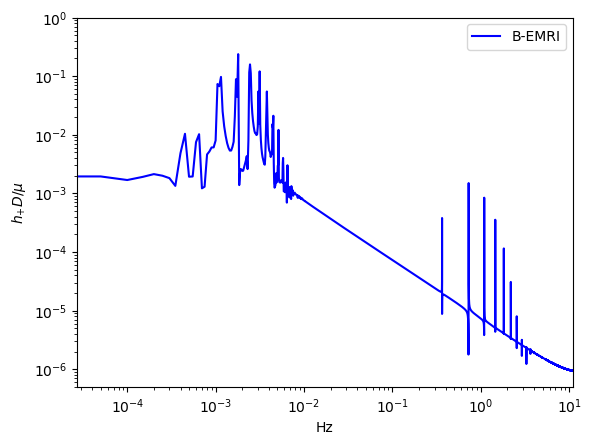}
\includegraphics[width=0.4 \textwidth]{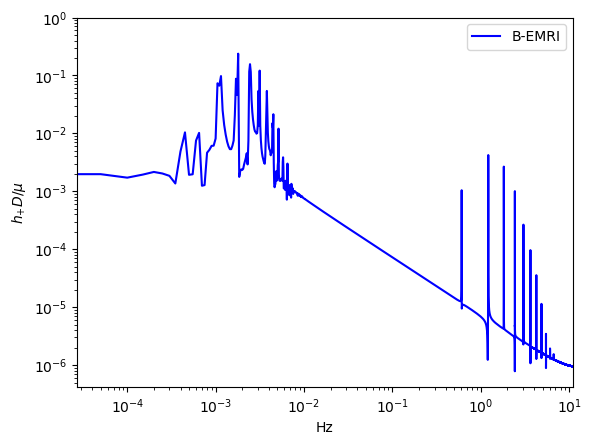}
\end{center}
\caption{Frequency spectrum of the GWs in Fig.\ref{waveformsmass}.
}\label{freqmass}
\end{figure*}

Fig.\ref{waveformstheta} displays the comparisons of the $h_{+}$ waveforms between EMRIs and B-EMRIs with different $\theta_0$, which shows that, as $\theta_0$ increases the amplitude of fluctuation
first decreases and then increases, especially when $\theta_0=0$ or $\theta_0=\pi$ the amplitude of fluctuation takes maximum value, i.e., when the rotation axis of the binary is parallel or
anti-parallel to the rotation axis of the SBH, the waveforms of B-EMRIs and EMRIs take the most sharp distinction.

In Fig.\ref{waveformsyimass} the $h_{+}$ waveforms of EMRIs and B-EMRIs with SBH weighted $M=10^8M_{\odot}$  are displayed. One sees that, since the central SBH is much heavier in this case, the
amplitude of fluctuation generated by binary composed of compact objects weighted $8M_{\odot}$ and $10M_{\odot}$ is very tiny. This is because for heavier BH with parameter selection $p(0)=10M$ the
cycle is much longer and frequency is much smaller. As displayed in Fig.\ref{waveformsmass}, like the $M=10^6M_{\odot}$ case as mass increases the amplitude of fluctuation increases.

\begin{figure*}
\begin{center}
\includegraphics[width=0.32 \textwidth]{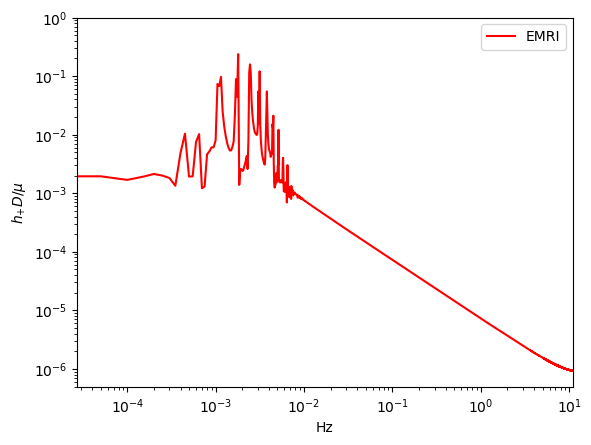}
\includegraphics[width=0.32 \textwidth]{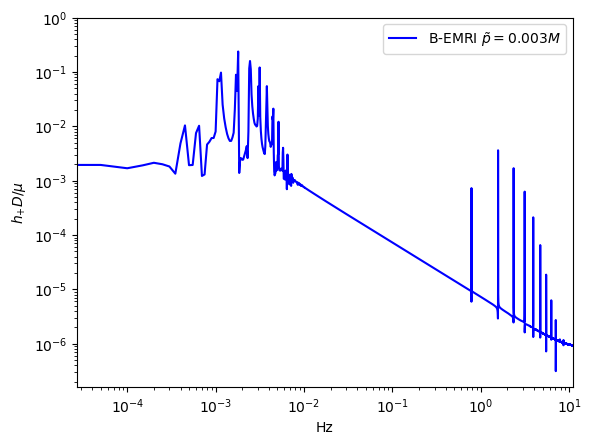}
\includegraphics[width=0.32 \textwidth]{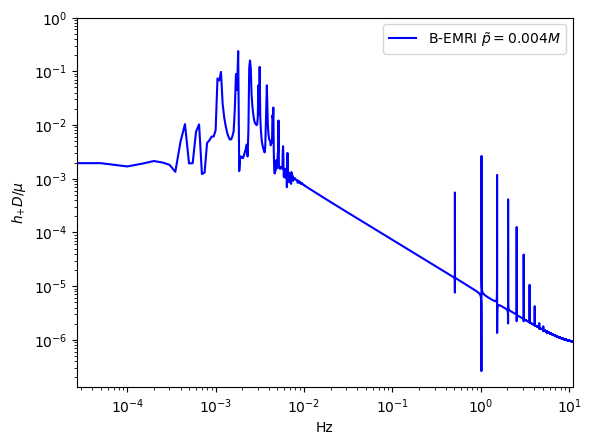}
\includegraphics[width=0.32 \textwidth]{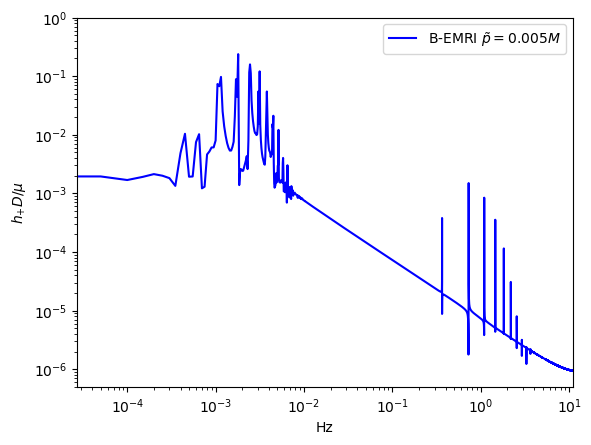}
\includegraphics[width=0.32 \textwidth]{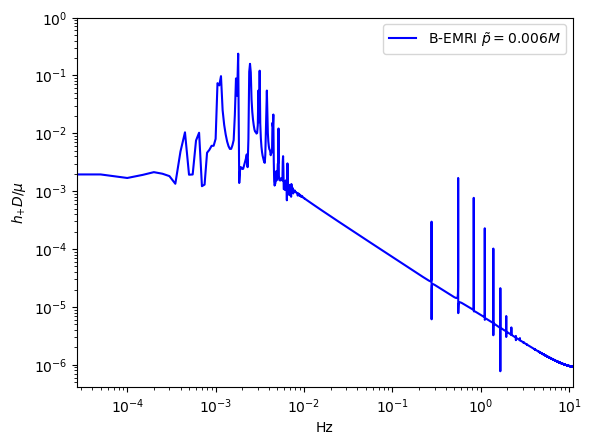}
\includegraphics[width=0.32 \textwidth]{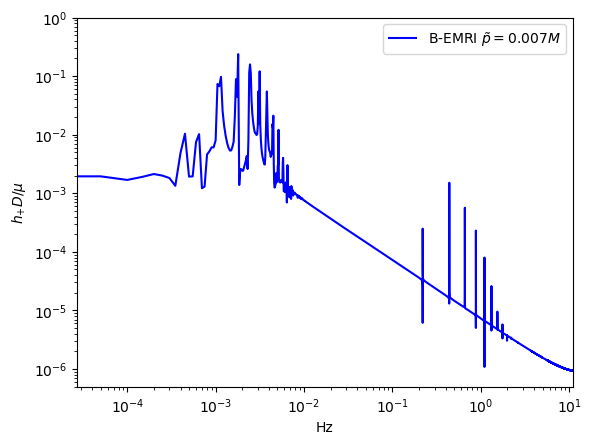}
\end{center}
\caption{Frequency spectrum of the GWs in Fig.\ref{waveformsp}.
}\label{freqp}
\end{figure*}

To compare the waveforms of B-EMRIs and EMRIs we analyse the waveforms in frequency domain. In Fig.\ref{freqmass} we present the frequency spectrum of the waveforms in Fig.\ref{waveformsmass}. One
sees that, in low frequency region the spectrum of B-EMRIs coincide with that of EMRIs. In high frequency region, only the B-EMRIs have non-vanishing amplitudes of gravitational signals. Further, as
mass increases, the amplitudes of signals increase, and the more and more higher-frequency signals emerge. This result agrees with the waveforms in Fig.\ref{waveformsmass}, where it shows as mass
increases the amplitude and frequency of the fluctuations become larger.  This is because, on the one hand, the amplitude of GW is proportional to the mass of the source, on the other hand, as mass
increases, the cycle decreases and the frequency increases. From the GW expression (\ref{quadoct}) we know, as frequency increases the amplitude of GW increases as well.

\begin{figure*}
\begin{center}
\includegraphics[width=0.32 \textwidth]{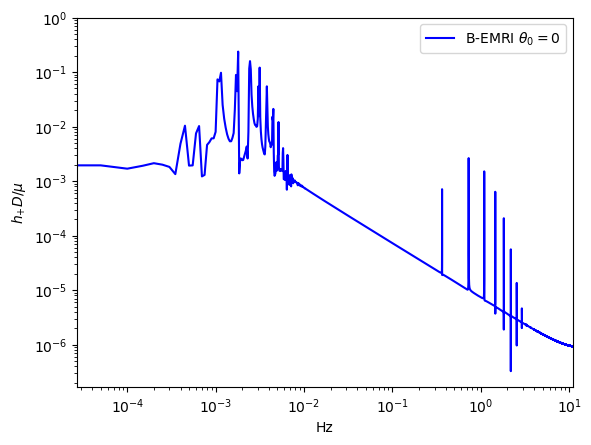}
\includegraphics[width=0.32 \textwidth]{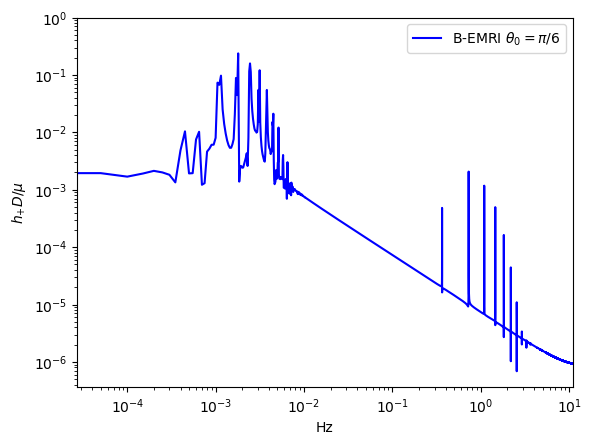}
\includegraphics[width=0.32 \textwidth]{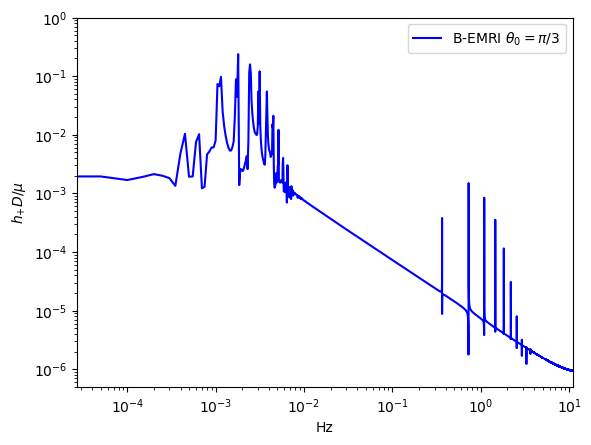}
\includegraphics[width=0.32 \textwidth]{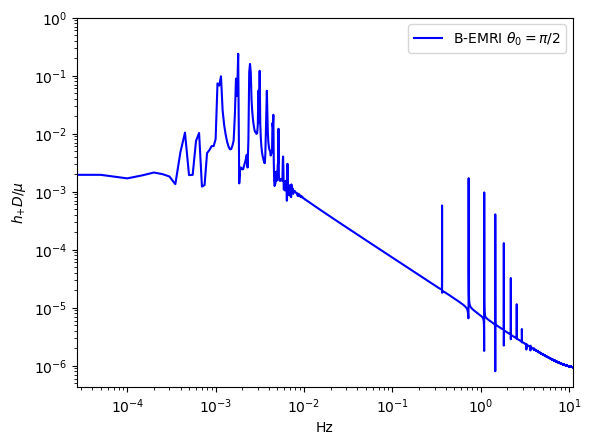}
\includegraphics[width=0.32 \textwidth]{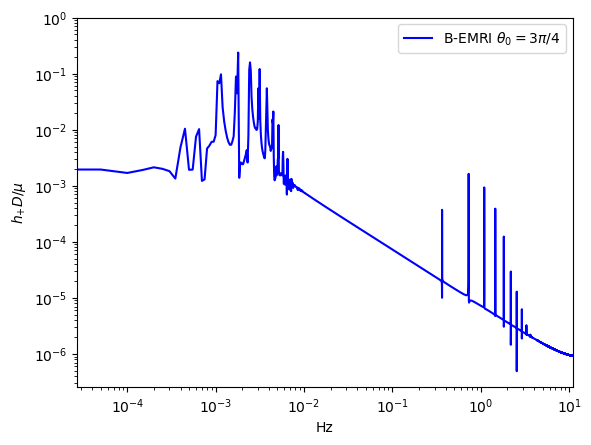}
\includegraphics[width=0.32 \textwidth]{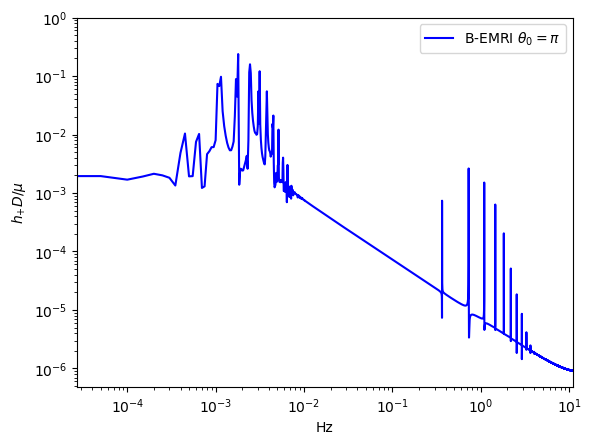}
\end{center}
-\caption{Frequency spectrum of the GWs in Fig.\ref{waveformstheta}. In  high frequency region the spectrum of EMRIs is definitely different from the spectrum of B-EMRIs. We can distinguish B-EMRIs
from EMRIs without question in frequency domain by using the data from space-based GW detectors.
}\label{freqtheta}
\end{figure*}

In Fig.\ref{freqp}, we give the frequency spectrum of the GWs in Fig.\ref{waveformsp}. We can see, as $\tilde{p}$ decreases, the amplitude of GW signal increases, and more and more high-frequency
signals emerge, and vice versa. This agrees with what the waveform shows. Since the smaller $\tilde{p}$ is, the shorter the cycle is, and the higher the frequency is, so the amplitude of the signal
increases.
In Fig.\ref{freqtheta} we give the frequency spectrum of the waveforms in Fig.\ref{waveformstheta} with distinct $\theta_0$. We can see, for distinct $\theta_0$, the frequency range is not
significantly different. The amplitude of the signals with different frequency first decreases and then increases. For $\theta_0=0$ or $\theta_0=\pi$, the amplitude of the signals takes the maximum
value, which agrees with the waveforms in Fig.\ref{waveformstheta}.

When GEM force is taken into account,  phase shift between the GW waveforms of B-EMRIs and that of EMRIs emerges, as shown in Fig.\ref{Binary-GEM}. It's sensitive and valuable to calculate  mismatch
of the two waveforms to quantify the distinction between them.
The inner product of two signals is defined as
\begin{eqnarray}
\langle x,h\rangle =  2 \int_0^{+\infty} \frac{\tilde{x}(f)\tilde{h}^*(f) + \tilde{x}^*(f)
\tilde{h}(f)}{S_h(f)} df,\label{innerproduct}
\end{eqnarray}
where $\tilde{x}(f)$ and $\tilde{h}(f)$ are the Fourier transformations of time-domain signals $x(t)$ and $h(t)$
 \begin{eqnarray}
\tilde{x}(f) = \int^\infty_{-\infty} x(t)e^{-i2\pi tf}dt,\quad\quad\quad\quad\tilde{h}(f) = \int^\infty_{-\infty} h(t)e^{-i2\pi tf}dt.
%\hat{h}_k = \sum_{j=0}^{j=M-1} y(t_i)e^{-i2\pi kj/M} ,
\end{eqnarray}
Here $*$ denotes the complex conjugate of a quantity, and $S_h(f)$ in (\ref{overlap}) is the one-sided noise power spectral density of LISA \cite{Robson:2018ifk}, which is significant for surveying
the types of sources that can be detected by the LISA mission.

With the inner product defined in (\ref{innerproduct}), the overlap of two signals is given by
\be
\mathcal{O}=\frac{\langle x,h\rangle}{\sqrt{\langle x,x\rangle}\sqrt{\langle h,h\rangle}}.\label{overlap}
\en
Mismatch of two GW signals is defined as
\be
\Delta=1-\mathcal{O},
\en
with which the indistinguishable criterion of two
signals is given by \cite{Baird:2012cu,Toubiana:2024car,Flanagan:1997kp,Chatziioannou:2017tdw,Mangiagli:2018kpu,Tan:2024utr,Mitman:2025tmj,
Zhao:2025sck,Lindblom:2008cm,Drummond:2023wqc,Yu:2023lml,Purrer:2014fza,Ohme:2011rm,Lynch:2021ogr}
\begin{equation}
\Delta\leq \frac{1}{2\rho^2},
\end{equation}
For LISA, the SNR is taken to be $\rho=20$\cite{Babak:2017tow}, from which we learn two signals are distinguishable when $\Delta>0.00125$. \textcolor{black}{In Fig.\ref{Binary-GEM}, we compare the gravitational waveforms of both the B-EMRI system without GEM force (left panel) and the standard EMRI system (right panel) against those of the B-EMRI system that includes GEM force.} Calculation shows for the two waveforms in the first
panel of Fig.\ref{Binary-GEM}  $\Delta=0.003921$, for the two waveforms in the second panel of Fig.\ref{Binary-GEM} $\Delta=0.003922$, which means when taking into account GEM force the waveforms of
B-EMRIs are distinguishable from the waveforms of B-EMRIs that does not take into account GEM force, and are distinguishable from the waveforms of EMRIs as well. \textcolor{black}{In Fig.\ref{Binary-GEM} the mass of the central SBH is set to be $10^8M_{\odot}$ . In Fig.\ref{Binary-GEM-106sun}, we set the mass of central SBH to be $4\times10^6M_{\odot}$ and compare the gravitational waveforms with/without GEM force for different eccentricities of internal orbits. Fig.\ref{Binary-GEM-106sun} shows that, the mismatch between the two types of waveforms accumulates with time. The larger the eccentricity of the inner orbit, the earlier the two waveforms begin to exhibit differences. For $\tilde{e}=0.2$, the amplitudes of the two waveforms become significantly different after about 30 days. For $\tilde{e}=0.5$, the amplitudes of the two waveforms become significantly different after about 25 days. For $\tilde{e}=0.8$, the amplitudes of the two waveforms become significantly different after about 18 days.  So as the observation time increases, the two waveforms ultimately become distinguishable.}

\section{Conclusion and Discussion \label{section5}}
In this paper we study the GWs emanated from B-EMRIs which is composed of a central SBH and binary revolving around it. B-EMRIs are quite popular in the universe, and remarkable objects for GW
detectors of next generation. We demonstrate that the waveform of B-EMRIs is distinctly different from that of EMRIs in frequency domain, and show that the mismatch can be effectively probed by the
space-based GW detectors such as LISA, Taiji, and Tianqin. Our results will improve template construction for B-EMRIs and shed light on detection of GWs from B-EMRIs.

 We solve the GW equations by taking into account the mass quadrupole, mass octupole and current quadrupole moments of the source. For the radiation reaction we take the hybrid 2nd order
 post-Newtonian approximation which is applicable to generic inclined-eccentric orbits. We plot the GW waveforms of B-EMRIs and compare the waveforms of B-EMRIs with that of EMRIs of single star.
 The results show that, the waveforms generated by single stellar-mass compact object share the same profile with the ones generated by binary compact objects, but the details of the waveforms of
 B-EMRIs exhibit fluctuations. The fluctuations are caused by internal motions of the binary. As masses of the compact objects of binary increase, the amplitudes and frequencies of the fluctuations
 increase. This is because, on the one hand, the amplitude of GW is proportional to the mass of the source. On the other hand, the increase of mass of gravitational source cause the decrease of
 cycle, so the frequency increases, which increases the amplitude further.  We fix other parameters and only adjust the semi-latus rectum of internal orbits $\tilde{p}$, the result shows that, as
 $\tilde{p}$ decreases, the amplitudes and frequencies of the fluctuations increase. This is because, as $\tilde{p}$ decreases, the cycle of internal motion decreases, the frequency increases, which
 causes the increase of amplitudes of GWs. We fix other parameters and adjust the angle $\theta_0$ between axes $z$ and $z_0$, the results show that, when $\theta_0=0$ or $\pi$ the amplitudes of GW
 fluctuations caused by binary internal motions are larger than the cases when $\theta_0$ takes other values, which means when $\theta_0=0$ or $\pi$ the GWs of B-EMRIs are most easily distinguished
 from that of EMRIs.

  For the binary on the equatorial plane of SBH, we further take in account GEM force and calculate the GWs of B-EMRIs. By comparing with both the waveforms of B-EMRIs that does not take into
  account GEM force and the waveforms of EMRIs, it's easy to find there is phase shift between the related two waveforms.

To compare the GWs of EMRIs and B-EMRIs further we  perform the frequency spectrum analysis. The results show that, at low-frequency region, the frequency spectrum of B-EMRIs coincide with that of
EMRIs. At high-frequency region, the frequency spectrum of B-EMRIs and EMRIs are different. As masses of the binary increase, more high-frequency signals emerge, and the amplitudes of the GW signals
increase, which agrees with the waveforms show us. We give the frequency spectrum of GWs generated by binary with distinct semi-latus rectum of internal orbits $\tilde{p}$, the results show that, as
$\tilde{p}$ decreases, more high-frequency signals emerge, and the amplitude of the GW signals increase, since smaller $\tilde{p}$ corresponds to shorter cycle and higher-frequency orbit motions,
higher-frequency orbit motions generate GWs with higher amplitudes. If we adjust the angle $\theta_0$ between axes $z$ and $z_0$, frequency spectrum show that when $\theta_0=0$ or $\pi$ the
amplitudes of GWs are larger than that of GWs when $\theta_0$ takes other values, this agrees with what the waveforms show us as well. So the GW signals of B-EMRIs can be clearly distinguished from
that of EMRIs through spectrum analysis, the GW signals generated by binary with larger mass, closer distance are more easily to be distinguished from that generated by single compact object.
Moreover, when rotation axis of the binary is parallel or anti-parallel to that of the central SBH, the GW signals of B-EMRIs are more easily to be distinguished from that of EMRIs. When taking into
account GEM force, we analyze the waveforms by calculating mismatch, calculation shows \textcolor{black}{mismatch increases with time. As time increases, mismatch accumulates, eventually} the gravitational waveforms of B-EMRIs \textcolor{black}{with GEM force} are distinguishable from  \textcolor{black}{those} of B-EMRIs  \textcolor{black} {without}
GEM force and from  \textcolor{black}{those} of EMRIs. \textcolor{black}{The larger the eccentricity of the inner orbit, the earlier the two waveforms become distinguishable.}

\begin{figure*}[h]
\begin{center}
\includegraphics[width=0.4 \textwidth]{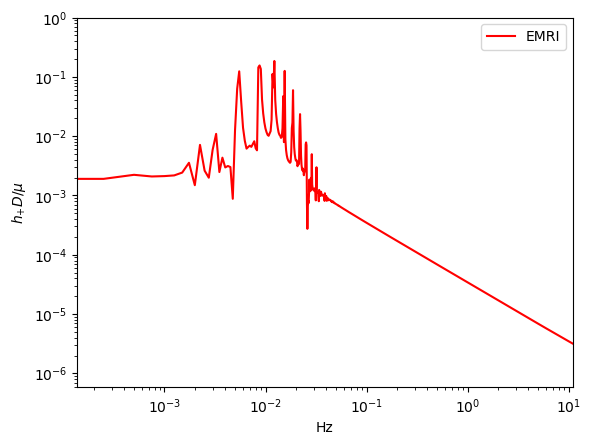}
\includegraphics[width=0.4 \textwidth]{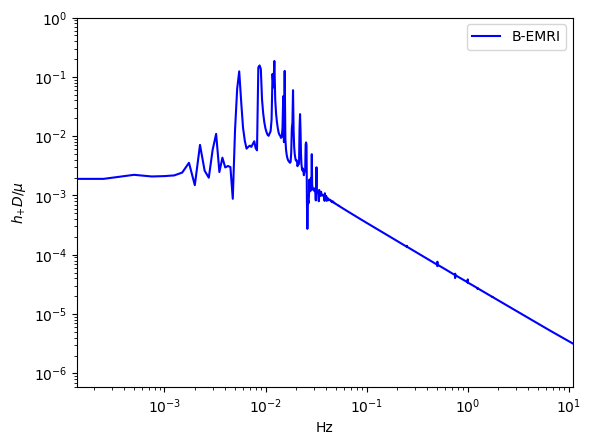}
\includegraphics[width=0.4 \textwidth]{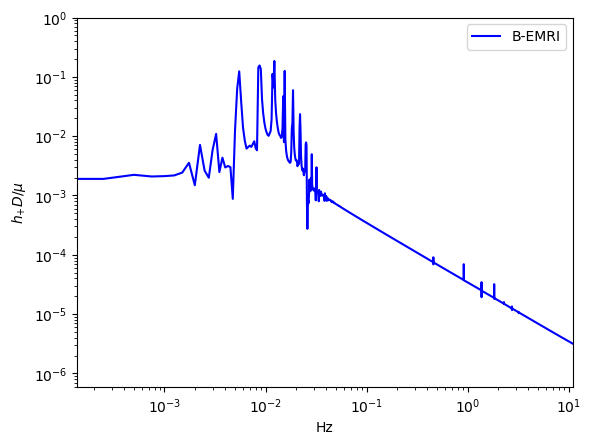}
\includegraphics[width=0.4 \textwidth]{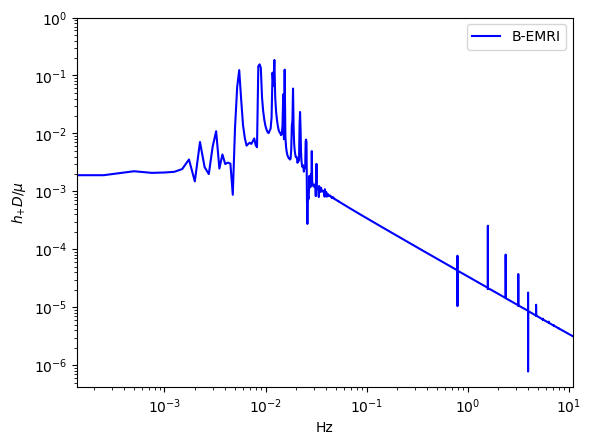}
\end{center}
\caption{Frequency spectrum of the GWs in Fig.\ref{waveformsyimass}.
}\label{freqyimass}
\end{figure*}

\section*{Appendix\label{section6}}
The $e$-dependent coefficients appear in Eq.(\ref{ELQdot}) are given by
\begin{eqnarray*}
g_1(e) &=& 1 + \frac{73}{24} e^2  + \frac{37}{96}e^4, \quad\quad\quad
g_2(e)= \frac{73}{12} + \frac{823}{24} e^2 + \frac{949}{32}e^4
+ \frac{491}{192}e^6 ,\
\nonumber \\
\nonumber \\
g_3(e) &=& \frac{1247}{336} + \frac{9181}{672} e^2 ,\quad\quad\quad
g_4(e)  = 4 + \frac{1375}{48} e^2  ,\
\nonumber \\
\nonumber \\
g_5(e) &=& \frac{44711}{9072} + \frac{172157}{2592} e^2 ,
\quad\quad
g_6(e) = \frac{33}{16} + \frac{359}{32} e^2 ,\quad\quad
g_9(e) = 1 + \frac{7}{8} e^2    ,
\nonumber \\
\nonumber \\
g_{10}^{a}(e)&=& \frac{61}{24} + \frac{63}{8}e^2   + \frac{95}{64}e^4 ,\
\quad\quad
g_{10}^{b}(e)= \frac{61}{8} + \frac{91}{4}e^2  + \frac{461}{64}e^4 .\
\nonumber \\
\nonumber \\
g_{11}(e) &=& \frac{1247}{336} + \frac{425}{336} e^2  ,\quad\quad\quad\quad
g_{12}(e) = 4 + \frac{97}{8} e^2 ,\
\nonumber \\
\nonumber \\
g_{13}(e) &=& \frac{44711}{9072} + \frac{302893}{6048} e^2 ,\quad\quad
g_{14}(e) = \frac{33}{16} + \frac{95}{16} e^2 ,\
\nonumber \\
\nonumber \\
g_{15}(e) &=& \frac{8191}{672} + \frac{48361}{1344} e^2 ,\quad\quad\quad
g_{16}(e) = \frac{417}{56} - \frac{37241}{672} e^2 .\
\end{eqnarray*}

\section*{Acknowledgment}
This work is supported by the National Key R$\&$D Program of China (no.2021YFC2203002), the National Natural Science Foundation of China Grants Nos. 12275106 and 12235019, and Natural Science
Foundation of Shandong Province Nos.ZR2023MA014.

\providecommand{\href}[2]{#2}\begingroup%\raggedright
\footnotesize\itemsep=0pt
\providecommand{\eprint}[2][]{\href{http://arxiv.org/abs/#2}{arXiv:#2}}

\end{document}